\documentclass[sigconf,nonacm]{acmart}
\AtBeginDocument{%
  \providecommand\BibTeX{{%
    \normalfont B\kern-0.5em{\scshape i\kern-0.25em b}\kern-0.8em\TeX}}}
\usepackage{hyperref}
\usepackage{pifont}
\usepackage{adjustbox}
\usepackage[ruled,linesnumbered,noend]{algorithm2e}
\usepackage{amsmath}
\usepackage{balance}
\usepackage{bbm}
\usepackage{booktabs}
\usepackage{color,colortbl}
\usepackage{cleveref}
\usepackage{dsfont}
\usepackage{enumitem}
\usepackage{float} 
\usepackage{subcaption} 
\usepackage{graphicx}
\usepackage{makecell}
\usepackage{mathtools}
\usepackage{multirow} 
\usepackage{pgfplots}
\usepackage{rotating}
\usepackage{subcaption}
\usepackage{tabularx}
\usepackage{threeparttable}
\usepackage{tikz}
\usepackage{tikz-qtree}
\usepackage{todonotes}
\usepackage{url}
\usepackage{utfsym}
\usepackage{varwidth}
\newtoggle{techreport}

\usepackage[dvipsnames, svgnames, x11names]{xcolor}

\usetikzlibrary{patterns}

\newcommand\vldbdoi{XX.XX/XXX.XX}
\newcommand\vldbpages{XXX-XXX}
\newcommand\vldbvolume{19}
\newcommand\vldbissue{5}
\newcommand\vldbyear{2025}
\newcommand\vldbauthors{\authors}
\newcommand\vldbtitle{\shorttitle} 
\newcommand\vldbavailabilityurl{URL_TO_YOUR_ARTIFACTS}
\newcommand\vldbpagestyle{plain} 

\newcommand{\sep}[1]{\textcolor{purple}{[Sep: #1]}}

\newcommand{\sys}{\textsc{HoldUp}\xspace}

\begin{document}

\newcommand{\Method}{{}}

\setlist[itemize]{leftmargin=2.0em}

\title{Semantic Data Processing with Holistic Data Understanding}
\author{Youran Sun$^{\dagger}$, Sepanta Zeighami$^{\dagger}$, Bhavya Chopra, Shreya Shankar, Aditya G. Parameswaran}
\affiliation{%
UC Berkeley\\
\{\url{youransun, zeighami, bhavyachopra, shreyashankar, adityagp}\} \url{@ berkeley.edu}
\country{}
\thanks{$^\dagger$Co-first author}
}
\begin{abstract}
Semantic operators have increasingly become integrated within data systems to enable processing data using Large Language Models (LLMs).
Despite significant recent effort in improving these operators, their accuracy is limited due to a critical flaw in their implementation:  \textit{lack of holistic data understanding}. In existing systems, semantic operators often process each data record independently using an LLM, without considering data context, only leveraging LLM's \textit{dataset-agnostic} interpretation of the user-provided task. However, natural language is imprecise, so a task can only
be accurately performed if it is correctly interpreted in the
context of the dataset.
For example, for classification and scoring tasks, which are typical semantic map tasks, the standard method of processing each record row by row yields inaccurate results in a wide range of datasets. 
We propose \sys, a new method for semantic data processing with holistic data understanding. \sys processes records jointly, leveraging cross-record relationships to correctly interpret the task within the data context. Enabling holistic data understanding, however, is challenging due to what we call \textit{LLM data understanding paradox}: while large representative data subsets are necessary to provide context, feeding long inputs to LLMs causes quality degradation due to well-known long-context issues. 
To resolve this paradox, we develop a novel clustering algorithm to identify the latent structure within the dataset through judicious use of LLMs, inspired by bagging. 
Using this approach as a primitive, we develop novel clustering-based classification and scoring methods to perform these two tasks with high accuracy. Experiments across 15 real-world datasets show that \sys consistently outperforms existing solutions, providing \textbf{\textit{up to 33\% higher accuracy for classification}} and \textbf{\textit{30\% higher accuracy for scoring and clustering}} tasks. 
\end{abstract}

\maketitle

\pagestyle{\vldbpagestyle}

\newcommand{\sun}[1]{\textcolor{teal}{[Youran: #1]}}

\section{Introduction}\label{sec:intro}

Many data systems~\cite{snowflake25, alloydb, databricks-llm, duckdb-llm, shankar2024docetl, patel2024lotus, liu2024declarative} now support \textit{semantic operators} that enable text data processing based on a user-specified natural language (NL) instruction using Large Language Models (LLMs). For example, a \textit{semantic map} applies an LLM-powered map operation to all record in a dataset based on an NL user instruction. 

\textbf{Lack of Holistic Data Understanding}. Despite significant recent effort in improving semantic operators~\cite{snowflake25, alloydb, databricks-llm, duckdb-llm, shankar2024docetl, patel2024lotus, liu2024declarative}, for many tasks \textit{\textbf{the accuracy of semantic operators is limited due to a critical flaw in their implementation: lack of holistic data understanding}.} Consider semantic maps, for example. Existing systems perform semantic map tasks by processing each record independently~\cite{patel2024lotus, shankar2024docetl, russo2025abacus}, without considering the task in the context of the rest of the dataset, similar to how a relational map would have been performed on a dataset. However, treating semantic operations similar to relational ones ignores the fact that \textit{\textbf{specification in natural language is imprecise by nature, so a task can only be accurately performed if it is correctly interpreted in the context of the dataset}}. Indeed, the importance of context when performing semantic tasks is well-established: humans (sometimes subconsciously) take overall context into account when making individual judgments~\cite{mozer2006context, zhu2007humans, hampton2006effects, barsalou1987instability}, as observed in crowdsourcing~\cite{zhuang2015debiasing, zhuang2015leveraging, scholer2013effect, mozer2010decontaminating}, and in the social sciences~\cite    {thaler2009nudge, schwarz1999self, schwarz2012context, tourangeau2000psychology}. We show that context matters for semantic data processing as well.


\begin{figure}[t]
    \centering
        \includegraphics[width=\columnwidth]{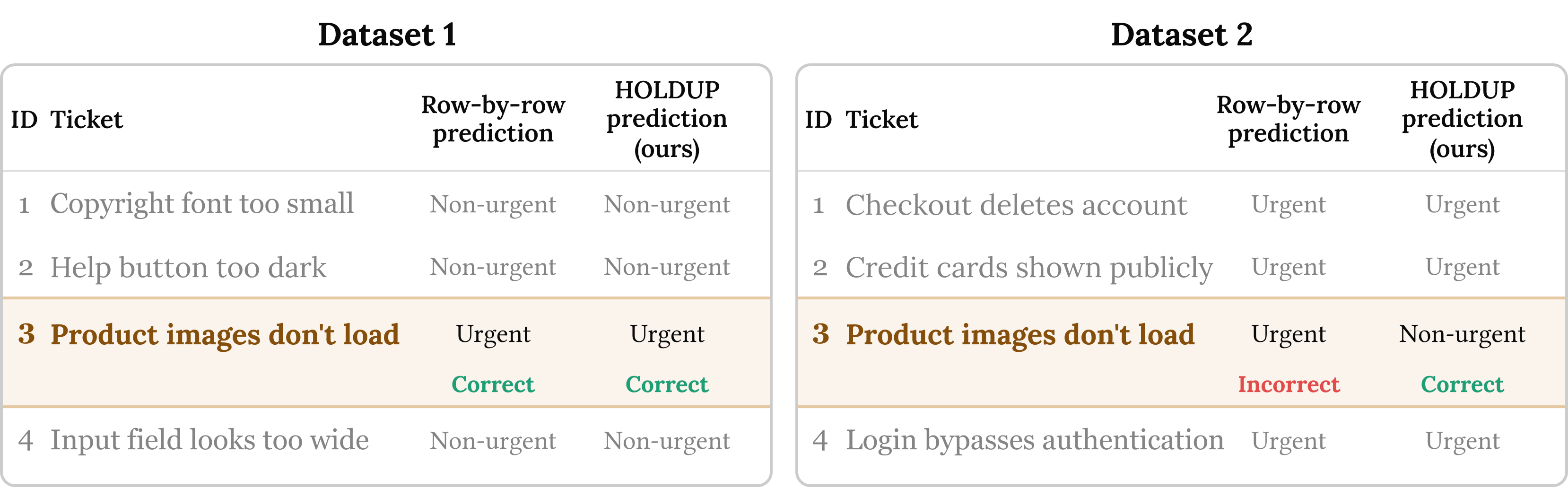}
        \caption{Classification example where correct label changes based on data context}
        \label{fig:classifications:example}
\end{figure}

For ease of exposition, we focus on classification and scoring, both typical map operations; we discuss other operations later. Both these tasks require assigning a label from a given set to each record in a dataset, class names in the former case and numerical scores in the latter. 
Existing systems'~\cite{patel2024lotus, shankar2024docetl, russo2025abacus} lack of holistic data understanding causes errors when the correct dataset-dependent interpretation of a task differs from LLM's dataset-agnostic one.
Fig.~\ref{fig:classifications:example} shows an example task of classifying support tickets for an e-commerce website based on urgency in two different datasets, where record \textcircled{3} is shared across datasets, but each dataset additionally contains other records.  In the context of Dataset 1, record \textcircled{3} would be considered an urgent issue, as it critically impacts functionality of the website, while other records are non-critical maintenance issues. Meanwhile, in the context of Dataset 2, record \textcircled{3} would be considered a non-urgent issue compared with other records that correspond to serious software and security issues.

LLMs' lack of holistic data understanding frequently causes errors in practice. Table~\ref{tab:classification-scoring-summary} shows a summary of our empirical results when performing classification and scoring tasks independently and row-by-row (Table~\ref{tab:classification-scoring-summary} first row), showing average accuracy across 10 real-world datasets for classification and 5 real-world datasets for scoring.  \textit{Our results show less than 50\% accuracy for scoring tasks and less than 70\% accuracy for classification tasks when processing data without holistic data understanding,} using top-of-the-line LLMs. 

\textbf{LLM Data Understanding Paradox.} Enabling holistic data understanding for semantic operators is challenging due to what we call \textbf{\textit{LLM data understanding paradox: to enable LLMs to understand the dataset, we need to provide large enough subsets of the dataset to the LLMs; however, providing large subsets of the data to an LLM worsens LLM quality}} as it increases input size, leading to well-known long-context issues~\cite{liu2024lost, wang2024beyond, bai2025longbench}. Because of this paradox, approaches that simply provide additional data subsets to the LLM as context during processing do not improve accuracy. Table~\ref{tab:classification-scoring-summary} rows 2-3 show that neither batching records together so LLM observes other records as context when making predictions (called Batch), nor asking an LLM to design a rubric, based on a data subset, to add to the prompt to provide data-dependent guidelines for the task  (called Rubric) improve the accuracy over independent row-by-row LLM calls on average.

\textbf{\sys}. We propose \sys, \underline{Hol}istic \underline{D}ata \underline{U}nderstanding for Semantic Data \underline{P}rocessing, a novel method that enables LLMs to holistically understand the dataset during semantic data processing, ensuring tasks are appropriately interpreted based on overall dataset context. 
Even for tasks specified as semantic maps, instead of processing records one by one, \sys jointly processes records while considering cross-record dependencies. We next discuss \sys for classification and scoring before other tasks. 

\begin{figure}[t]
    \centering
        \includegraphics[width=\columnwidth]{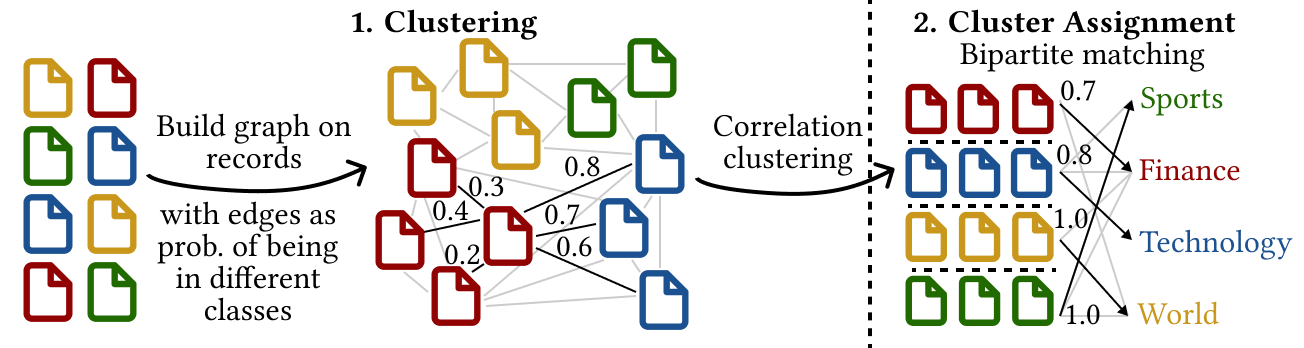}
        \caption{\sys workflow for multi-class classification}
        \label{fig:overview:classification}
\end{figure}

\textbf{Classification and Scoring with \sys.} Our key insight to enable holistic data understanding is to turn row-by-row classification into what we call \textit{clustering-based classification}. Clustering provides a natural way to identify related records and assign them to similar labels by holistically considering cross-record relationships. 
Unfortunately, this approach as specified does not work. First, accurately clustering the records is itself  difficult, again due to the LLM data understanding paradox---simply asking an LLM to perform clustering on the dataset has low accuracy. Second, even if we accurately identify the clusters, deciding which label each cluster represents is challenging, especially when there are many similar classes. Finally, cost can be an issue: clustering requires considering cross-record relationships which can be costly. We present novel clustering-based classification and scoring solutions that addresses these three challenges, discussed next. 

\textit{Clustering.} \sys clusters records into groups that belong to the same label. As mentioned earlier, doing so accurately is challenging because it requires reasoning about which records belong to the same label, based on the task instruction and the dataset. Simply using embedding-based clustering (e.g., using K-means on record embeddings) fails because it does not take the task instruction into account and because embeddings do not provide a fine-grained representation of the data. Meanwhile, using LLMs to perform clustering runs into the LLM data understanding paradox: it requires providing a large enough data subset to an LLM, which yields inaccurate results due to long context issues. 

We propose a novel clustering algorithm that addresses these challenges by (1)  using LLMs only for sub-tasks within clustering that they can perform accurately, and (2) developing novel methods to obtain accurate responses from LLMs to resolve the LLM data understanding paradox. We propose a graph-based clustering approach, where we construct a graph with nodes as records and edge weights as the probability of two records having different labels, estimated by an LLM. 
To accurately estimate this probability for two records, inspired by bagging~\cite{breiman1996bagging}, we repeatedly provide the records together with random subsets of the dataset to an LLM and ask it to output whether the records should belong to the same class, averaging the output across samples---repeated sampling improves  accuracy of probability estimation  while additional data context ensures estimates are informed by the dataset as whole. We then use correlation clustering~\cite{bansal2004correlation, gionis2007clustering} to cluster this graph. 
Fig.~\ref{fig:overview:classification} (b) shows an example, where edge weights are obtained using an LLM. 

\begin{table}[t]
    \small
    \hspace{-0.5cm}
    \begin{minipage}{0.55\columnwidth}
        \setlength{\tabcolsep}{2pt}
        \begin{tabular}{ccccc}
        \toprule
        & \multicolumn{2}{c}{\textbf{Classification}} & \multicolumn{2}{c}{\textbf{Scoring}} \\
        \cmidrule(lr){2-3} \cmidrule(lr){4-5}
        \textbf{Method}
        & \textbf{\makecell{Avg.\\acc.} }
        & \textbf{\makecell{Max. $\Delta$}}
        & \textbf{\makecell{Avg.\\acc.} }
        & \textbf{\makecell{Max. $\Delta$}} \\
        \midrule
        \textbf{\makecell{Row by\\Row}}  & 0.66 & $\uparrow$ 0.33 & 0.48 & $\uparrow$ 0.28 \\
        \textbf{Batch} & 0.43 & $\uparrow$ 0.57 & 0.46 & $\uparrow$ 0.33 \\
        \textbf{Rubric} & 0.59 & $\uparrow$ 0.36 & 0.50 & $\uparrow$ 0.29 \\
        \textbf{\sys} & \textbf{0.80} & - & \textbf{0.64} & - \\
        \bottomrule
        \end{tabular}
        \subcaption{Classification and scoring results.}
        \label{tab:classification-scoring-summary}
    \end{minipage}
    \hfill
    \begin{minipage}{0.44\columnwidth}
        \begin{tabular}{ccc}
        \toprule
        \textbf{Method} & \textbf{\makecell{Avg.\\acc.}}
        & \textbf{Max. $\Delta$}   \\  
        \midrule
        \textbf{\makecell{K-means+\\Embed.}} & 0.45 & $\uparrow$ 0.37 \\
        \textbf{Lotus} & 0.50 & $\uparrow$ 0.30 \\
        \textbf{\sys} & \textbf{0.70} & -  \\  
        \bottomrule
        \end{tabular}
        \subcaption{Clustering results.}
        \label{tab:clustering-summary}
    \end{minipage}
    \caption{Summary of \sys's performance across tasks. Avg. acc. is average accuracy and Max. $\Delta$ shows \sys's maximum accuracy improvement across all datasets in Sec.~\ref{sec:experiments}.}
    \label{tab:summary}
\end{table}


\textit{Cluster Assignment.} \sys then assigns each cluster to a label. This step also requires jointly considering all clusters, especially when there are many clusters. Simply asking an LLM to independently assign a label to each cluster causes errors for similar reasons as when labeling each record. Without a holistic understanding of all clusters, the LLM cannot interpret the task within the data context, while providing additional clusters as context yield inaccurate results due to the LLM data understanding paradox.  

To accurately assign labels to clusters, we use insights similar to our clustering algorithm, where we (1) jointly perform cluster assignment across clusters to take cross-cluster dependencies into account, (2) use LLMs for sub-tasks within cluster assignment that they can perform accurately, and, differently from our clustering algorithm, (3) additionally utilize task-specific characteristics to improve accuracy. For classification, we construct a bipartite graph between clusters and class names, where each edge weight reflects the probability---determined by an LLM---that a given cluster corresponds to a given class. We then compute a maximum-weight perfect matching to assign a unique label to each cluster. Fig.~\ref{fig:overview:classification} shows an example of this bipartite graph and the cluster assignment output. For scoring, we exploit the ordinal nature of scores to ensure accurate label assignment. Specifically, we use LLMs to decide pairwise orders between clusters---i.e., whether a cluster should be scored higher than another. We use these pairwise comparisons to sort the clusters while minimizing the number of pairwise relationships violated using an Integer Linear Program and assign a score to each cluster based on this ordering.

\textit{Cost Optimizations}. Finally, our clustering algorithm can be expensive---it requires obtaining probabilities of whether every pair of records belong to the same class. We propose multiple optimizations to reduce the overall cost. First, we only use our clustering algorithm for records that are difficult to label. Unambiguous records whose label can be accurately determined without data context are labeled through row-by-row processing---we use a model cascade approach~\cite{zeighami2025cut, patel2024lotus} to decide how each record should be processed. Second, for records processed with our clustering-based method, we perform clustering in constant size batches, which ensures cost remains linear in data size. Finally, we use a statistical model to decide how many samples are needed for accurate clustering, avoiding unnecessary LLM calls.

\textbf{Other tasks with \sys}. Our clustering algorithm can be used to perform clustering tasks as well, and semantic filter and joins are binary classification tasks (the latter on the cross product of two datasets), so that our classification algorithm is applicable to those tasks.  We believe our insights are also applicable to open-ended map operations (e.g., extraction and summarization), where holistic data understanding can improve task interpretation as well as consistency of outputs across records---we leave a more detailed study to future work.

\textbf{Contributions}. Our contributions are as follows:
\begin{itemize}
    \item We show that an important subset of real-world tasks require holistic data understanding and are not well-supported by existing semantic operator implementations,
    \item We propose a novel operator, \sys, that performs classification, scoring and clustering with high accuracy,
    \item We develop a novel clustering algorithm and clustering-based classification and scoring methods; and
    \item We present extensive experimental results on 15 real-world datasets, showing \textbf{\textit{\sys outperforms existing solutions, providing on average 14\% and up to 33\% higher accuracy for classification across datasets}} and up to \textit{\textbf{30\% for scoring and clustering}} with comparable costs. 

\end{itemize}

\section{Background and Overview}
\label{sec:background}
Our goal is to accurately perform semantic data processing by holistically understanding the dataset.  For ease of presentation, we first present our approach for classification, a common use-case for semantic data processing~\cite{shankar2024docetl, patel2024lotus, russo2025abacus, zeighami2025cut}, which includes filtering as a special case. We formalize the classification problem in Sec.~\ref{sec:classification:setup}, and overview our solution in Sec.~\ref{sec:classification:overview}, with details presented in Secs.~\ref{sec:class-method}-\ref{sec:class-extend}. We present extensions to other tasks, including scoring, clustering and joins in Sec.~\ref{sec:others}. 

\vspace{-0.05cm}
\subsection{Classification Setup and Background}\label{sec:classification:setup}
\textbf{Classification Task}.
We are given a dataset, $\mathcal{D}$, containing $n$ records, a natural language statement \texttt{P} specifying the \textit{classification instruction} (e.g., ``assign a topic to the news article'' for a dataset $\mathcal{D}$ of news articles) and a set of $k$ \textit{class names} $L=\{\ell_1, \ell_2, \cdots, \ell_k\}$ (e.g., ``world news'', ``sports''). The goal is to return a  \textit{prediction set} $\hat{Y}=\{\hat{y}_{1}, \hat{y}_{2}, \cdots, \hat{y}_{n}\}$, where $\hat{y}_{i}\in L$ is the \textit{predicted class} for the $i$-record in $\mathcal{D}$, with high \textit{classification accuracy}, $\mathcal{A}$, defined as
\vspace{-0.05cm}
\begin{align}\label{eq:classification_acc}
\mathcal{A}(Y, \hat{Y}) = \frac{1}{n}\sum_{i=1}^n \mathds{I}[\hat{y}_{i} = y_{i}],
\end{align}
where $Y=\{y_{1}, y_{2}, \cdots, y_{n}\}$ and $y_{i}$ is the \textit{ground-truth class} for the $i$-th record in $\mathcal{D}$, $i\in [n]$. 

\textbf{LLMs}. We predict the class records belong to using LLMs. We use $\mathcal{M}(x)$ to denote an LLM call with input $x$ and $\mathcal{M}(x_1,..., x_k)$ to refer to an LLM call with a prompt containing a concatenation of $x_1,..., x_k$. For an LLM call, $\mathcal{M}(x)$, we obtain the model's confidence in its output by $\mathscr{S}_\mathcal{M}(x)$, obtained from model log probabilities. We assume access to two different LLMs, a cheaper and potentially less accurate LLM $\mathcal{M}_c$, and a more expensive but more accurate LLM, $\mathcal{M}_e$; $\mathcal{M}(x)$ generically specifies an LLM call implemented with either $\mathcal{M}_e$ or $\mathcal{M}_c$ (our techniques can be extended to additional LLMs but we focus here on two for ease of exposition). We use the cheaper model to reduce cost when possible, using a model cascade framework, discussed next. 

\if 
We predict the class records belong to using LLMs. Similar to prior work~\cite{patel2024lotus, zeighami2025cut}, we assume access to two different LLMs, a cheaper and potentially less accurate LLM, $\mathcal{M}_c$ and a more expensive but more accurate LLM, $\mathcal{M}_e$---we use the cheaper model when possible to reduce cost. 
We use $\mathcal{M}_e(x)$ and $\mathcal{M}_c(x)$  to denote an LLM call with either of the two models with an input $x$, and use $\mathcal{M}(x)$ to generically specify an LLM call that could be implemented with either $\mathcal{M}_e(x)$ or $\mathcal{M}_c(x)$. We use $\mathcal{M}(x_1,..., x_k)$ to refer to an LLM call with a prompt that contains a concatenation of $x_1,..., x_k$. For an LLM call, $\mathcal{M}(x)$, we obtain the model's confidence in its output by $\mathscr{S}_\mathcal{M}(x)$. We use output log probabilities to obtain $\mathscr{S}_\mathcal{M}(x)$, similar to prior work \cite{zeighami2025cut, patel2024lotus}. 
\fi

\textbf{Model Cascade.} Similar to prior work~\cite{patel2024lotus, zeighami2025cut} we use model cascades to reduce cost, where for each record, we choose between using a cheap but inaccurate method, called a \textit{proxy}, or an expensive but accurate method, called an \textit{oracle}, to process the record.
We route records to proxies in order to reduce cost when possible while providing accuracy similar to that of the oracle. The decision of whether to use a proxy for a record is often based on proxy's confidence score (e.g., obtained from  model log-probabilities). A proxy is used to process a record if it's confidence for that record is above a threshold. The threshold is determined upfront through comparing proxy and oracle responses on a sample~\cite{patel2024lotus, zeighami2025cut}. 

\textbf{Cost Model}. We measure the cost of performing classification as the monetary cost of using LLMs; computational cost of non-LLM components is negligible. For an algorithm, $Alg$, we denote by $\texttt{cost}(Alg)$ the total monetary costs it incurs by using $\mathcal{M}_e$ and $\mathcal{M}_c$.

\textbf{Problem Definition}. 
Our goal is to perform classification with high accuracy given a monetary budget $\widetilde{\mathcal{C}}$ (with $\widetilde{\mathcal{C}}=\infty$ denoting no budget limit), 
that is, to design an algorithm, $Alg(\mathcal{D}, \texttt{P},  L)$, that given a dataset $\mathcal{D}$, classification instruction \texttt{P}, and class names $L$, satisfies $\texttt{cost}(Alg(\mathcal{D}, \texttt{P},  L))\leq \widetilde{\mathcal{C}}$, and returns a prediction set $\hat{y}$ with maximum classification accuracy.   

\textbf{Existing Solutions.} 
To solve the classification problem, existing data systems use a semantic map operator~\cite{snowflake25, alloydb, databricks-llm, duckdb-llm, shankar2024docetl, patel2024lotus, liu2024declarative}, by calling an LLM independently on every row to perform classification.  Specifically, $\mathcal{M}(\texttt{P}, t)$ is invoked for every $t\in \mathcal{D}$  independently, generating the prediction set $\hat{Y}=\{\mathcal{M}(\texttt{P}, t); t\in \mathcal{D}\}$, where $\mathcal{M}(\texttt{P}, t)$ denotes providing the instruction $\texttt{P}$ and a tuple $t\in \mathcal{D}$ to an LLM and asking it to perform classification. To decide which model, $\mathcal{M}_e$ or $\mathcal{M}_c$, should be used to implement the LLM call $\mathcal{M}$, some approaches leverage a model cascade solution~\cite{zeighami2025cut, patel2024lotus} to reduce cost, but still apply each model independently to each row. 

Such an approach yields inaccurate results when classification requires interpreting the task based on the dataset context. In practice, class boundaries often overlap and the correct class for a record depends on other records. Processing every record row-by-row ignores such dependencies, leading to poor accuracy.

\begin{figure}
\vspace{-0.7cm}
    \centering
    \includegraphics[width=1.0\columnwidth]{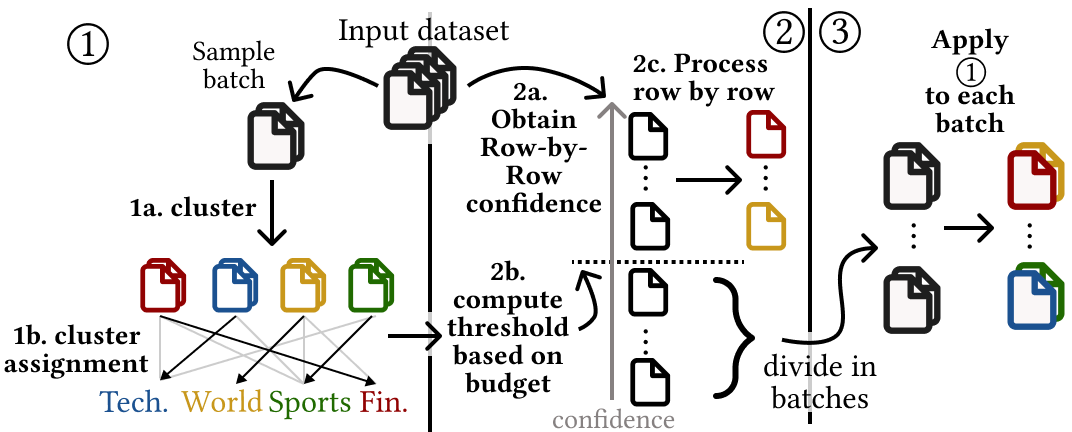}
    \caption{Classification with \sys}
    \label{fig:classification-framework}
\end{figure}

\begin{algorithm}
\footnotesize
    \caption{$\texttt{\sys}_{\text{classification}}(\mathcal{D}, \texttt{P}, k, L)$}
    \label{alg:classification-framework}

    \SetKwInOut{Input}{Input}
    \SetKwInOut{Output}{Output}
    \SetKwComment{tcp}{\textcolor{gray}{$\triangleright$}\ }{}
    \newcommand{\mytcp}[1]{\tcp{\textcolor{gray}{#1}}}
    \newcommand{\mytcpstar}[1]{\tcp*{\textcolor{gray}{#1}}}
    \SetCommentSty{textnormal}
    \SetKwProg{Fn}{Function}{}{}

    \Input{Dataset $\mathcal{D}$, task prompt \texttt{P}, integer $k$, class labels $L$} 
    \Output{Predicted classes for records in $\mathcal{D}$}

  
    \mytcp{\textcircled{1} Clustering-based classification on a sample batch $D_0$}
    $D_0 \gets$ a sample batch from $\mathcal{D}$;

    $\hat{Y}_{D_0}, \mathcal{C}_0 \gets \texttt{CbClassification}(D_0, \texttt{P}, k, L)$\;

    \mytcp{\textcircled{2} Model cascade with row-by-row processing}

    $\hat{Y}_{D_R}, D_R \gets \texttt{PredictWithCascade}(\mathcal{D}, \texttt{P}, k, L, D_0, \mathcal{C}_0)$
    



    
    \mytcp{\textcircled{3} Clustering-based classification in batches}
    $D_{X, 1}, \cdots, D_{X, n_B} \gets$ batches from $ \mathcal{D}\setminus D_0 \setminus D_R$\;

    $\hat{Y} \gets \hat{Y}_{D_0} \cup \hat{Y}_{D_R}$
    
    \For{$i \gets 1$ \KwTo $n_B$}{
        $\hat{Y}_{D_{X, i}}, \mathcal{C}_{X, i} \gets \texttt{CbClassification}(D_{X, i}, \texttt{P}, k, L)$\;
        $\hat{Y} \gets \hat{Y} \cup \hat{Y}_{X, i}$\;
    }

    \Return{$\hat{Y}$}

\vspace{0.5em}
\Fn{\texttt{CbClassification}$(D, \texttt{P}, k, L)$}{

    \Input{Dataset $D$, task prompt \texttt{P}, integer $k$, labels $L$}
    \Output{Predicted classes for records in $D$}
  
    $\hat{C} \gets \texttt{Cluster}(D, \texttt{P}, k)$
    \mytcpstar{$\hat{C} = \{\hat{C}_1, \hat{C}_2, \cdots, \hat{C}_k\}$}

    $\hat{Y} \gets \texttt{Assign}(\hat{C}, \texttt{P}, k, L)$\;

    $\mathcal{C} \gets$ total cost of \texttt{Cluster} and \texttt{Assign}\;
    \Return{$\hat{Y}_D, \mathcal{C}$}
}

\end{algorithm}

\vspace{-0.3cm}
\subsection{Classification with \sys}\label{sec:classification:overview}
The key insight in \sys is that the correct interpretation of the classification task and class names depends on all the records in the dataset.  Thus, rather than performing classification independently for each record, \sys processes records together while holistically considering cross-record relationships. 

\textbf{Overview}.  \sys uses a novel \textit{clustering-based classification} approach: this approach first clusters records to identify which ones should belong to the same class, then assigns classes to each cluster. The clustering step holistically considers cross-record relationships, enabling this approach to achieve high classification accuracy. However, clustering-based classification can be expensive, as it requires iterating over the data multiple times to evaluate which records should belong to the same class. On the other hand, row-by-row classification, as used by existing systems has low cost but can be inaccurate. To ensure \sys stays within the user's monetary budget, we combine both approaches following a model cascades-like framework, using clustering-based classification (as ``oracle'') on smaller subsets of the data that are difficult to classify, while using row-by-row classification (as ``proxy'') for the rest. 



Fig.~\ref{fig:classification-framework} illustrates \sys at a high level. \textcircled{1} \textbf{Initial clustering-based classification.} we sample a subset of the data, cluster these records and assign a class name to each cluster. \textcircled{2} \textbf{Clustering-informed row-by-row classification} for easy cases. We then use model cascades to decide a subset of the remaining records that can be processed using row-by-row classification, considering our clustering-based approach as the oracle and row-by-row classification with different models as proxies. We obtain LLM confidence scores for row-by-row classification and decide a threshold on the score based on cost-accuracy trade-offs from Step \textcircled{1} and the cost budget $\widetilde{C}$; records with score above this threshold are processed via row-by-row classification. 
\textcircled{3} \textbf{Clustering-based classification} for hard cases. For the rest of the data, we use our clustering-based method to perform classification. We perform clustering in constant-size batches to avoid a quadratic number of LLM calls, applying clustering-based classification independently to each batch that yields  cost linear in data size.

\textbf{Algorithm}. Algorithm~\ref{alg:classification-framework} presents this process.
We first sample a batch $D_0$ from $\mathcal{D}$ (line 1) and apply our clustering-based classification method, \texttt{CbClassification}, to $D_0$. The algorithm first partitions the records in $D_0$ into $k$ clusters based on which records are expected to belong to the same class, and then assigns each cluster to a distinct ground-truth class name (defined in Sec.~\ref{sec:class-method}), obtaining classification result $\hat{Y}_{D_0}$ as well as classification cost $\mathcal{C}_0$ (line 2).
Then, we use our cascade approach \texttt{PredictWithCascade} (defined in Sec.~\ref{sec:class-extend}), which classifies a subset $D_R$ of $\mathcal{D}$ using row-by-row classification (line 3). 
Finally, the remaining records are partitioned into batches (line 4), and each batch is independently classified using the clustering-based method (lines 5-8), again applying \texttt{CbClassification}.

\textbf{Roadmap}. We discuss how we perform clustering-based classification, \texttt{CbClassification}, in Sec.~\ref{sec:class-method} and present details of how we use a model cascade approach to stay within the budgetary constraint, \texttt{PredictWithCascade}, in Sec.~\ref{sec:class-extend}. Finally, Sec.~\ref{sec:others} shows how the approach can be extended to other tasks.

\vspace{-0.2cm}
\section{Clustering-Based Classification}
\label{sec:class-method}
Next, we discuss \texttt{CbClassification}, our clustering-based classification method that classifies a given set of data using clustering to enable holistic dataset understanding. 
Given an input dataset $D\subseteq \mathcal{D}$ (e.g., in step \textcircled{1} of Alg.~\ref{alg:classification-framework}, $D$ is set to $D_0$),  as shown in Alg~\ref{alg:classification-framework}, \texttt{CbClassification} first clusters records in $D$ into a set of $k$ clusters according to the classification instruction \texttt{P} (line 11, Alg~\ref{alg:classification-framework}) and then maps each cluster to a class name via bipartite matching (line 12, Alg~\ref{alg:classification-framework}). Without loss of generality, we let $D=\{t_1, ..., t_B\}\subseteq \mathcal{D}$ for an integer $B=|D|$. 
We discuss the clustering and cluster assignment steps in Secs.~\ref{sec:clustering} and~\ref{sec:matching}, respectively. 

\vspace{-0.2cm}
\subsection{Clustering Algorithm}
\label{sec:clustering}
Identifying accurate clusters of records that should belong to the same class at low cost is challenging. Naively clustering records using embeddings (e.g., by running K-means) yields inaccurate results because embeddings (1) don't provide a fine-grained representation of the data, and (2) don't take $\texttt{P}$, the classification instruction, into account causing errors when semantically dissimilar records can belong to the same classes (e.g., vastly different articles about ``African history'' and ``global warming'' may both be classified as ``world news''). Meanwhile, providing many data records to an LLM to cluster yields inaccurate results due to the \textit{LLM data understanding paradox}: (1) to obtain accurate answers we must provide a large subset of the data to the LLM, but (2) the LLM's accuracy degrades as more data is provided as input. In the case of clustering, a small data sample may not contain enough representative records to allow for accurate clustering, while providing more records leads to a longer input that LLMs are less accurate at processing~\cite{liu2024lost, wang2024beyond, bai2025longbench}.  

\textbf{Overview.} We develop a novel clustering method that addresses these shortcomings by (1) using LLMs only for subtasks they can perform accurately, rather than the clustering task itself, and (2) developing novel methods to obtain accurate responses from LLMs on large datasets to resolve the LLM data understanding paradox. 

We specifically propose a graph-based clustering approach to solve the problem. We represent the input dataset $D$,  with a weighted complete graph $G=(D, E, W)$, where each vertex is a record in $D$,  $E$ is the set of all record pairs (since $G$ is a complete graph) and $W$ is a weight matrix where $W[i, j]$ is the edge weight between the $i$-th and $j$-th record. We use $W[i, j]$ to denote the probability that the records $t_i$ and $t_j$ belong to different classes based on the classification instruction \texttt{P}. Our goal is to partition $G$ into $k$ clusters, $\hat{C}_1, ..., \hat{C}_k$ such that all records in $\hat{C}_i$ belong to the same class. 

This formulation reduces the clustering problem to one of accurately identifying the probability that two records belong to the same class. If we can identify the edge weights accurately, clustering the records is simple and can be done using any graph clustering method (we use correlation clustering~\cite{gionis2007clustering, bansal2004correlation}). For example, if we know the ground-truth classes, $y_i$ for all $i \in [|D|]$, we can let $W[i, j]=\mathds{I}[y_{i}\neq y_{j}]$ and trivially cluster $G$ by taking components that are connected with 0-weight edges. Nonetheless, obtaining accurate edge weights remains challenging. Similar to the original classification problem, whether two records should belong to different classes still requires holistic data understanding, which again leads to the LLM data understanding paradox, now, because to accurately decide whether two records belong to different classes we must provide additional data as context to the LLM, but LLM accuracy degrades as it is provided with more data.  

We propose a bagging-inspired~\cite{breiman1996bagging} approach to resolve this paradox. 
To obtain edge weights for a batch of data $D\subseteq \mathcal{D}$, we repeatedly provide subsets of this batch to the LLM to identify which records should belong to the same class. We aggregate LLM responses across the subsets to decide the probability that two records belong to a different class, which are then used as edge weights. 

We use this approach to obtain edge weights and cluster the data, doing so iteratively while alternating between obtaining edge weights and clustering. 
Specifically, our clustering algorithm iterates between (1) querying the LLM on a new sample to refine edge weights (discussed in Sec.~\ref{sec:clustering:edge_weights}) and (2) clustering the data based on current edge weights (discussed in Sec.~\ref{sec:clustering:corr}). We terminate when the clustering result stabilizes, i.e., is expected not to change with additional samples, formalized in Sec.~\ref{sec:clustering:corr}.

\textbf{Algorithm.} Alg.~\ref{alg:semantic-cluster} shows this process. We iteratively refine edge weights by taking new samples (line 4), then apply correlation clustering and decide whether to terminate (line 5). The algorithm terminates if we decide clusters have stabilized (i.e., further edge weight refinement will not impact the clustering outcome) or if we reach a maximum number of iterations $m_{\max}$ (line 2). We discuss \texttt{UpdateEdgeWeights} which computes the edge weights in Sec.~\ref{sec:clustering:edge_weights} and the correlation clustering and termination decision in Sec.~\ref{sec:clustering:corr}.

\begin{algorithm}[t]
\footnotesize
\caption{\texttt{Cluster}($D, \texttt{P}, k$)}
\label{alg:semantic-cluster}

\SetKwInOut{Input}{Input}
\SetKwInOut{Output}{Output}
\SetKwComment{tcp}{\textcolor{gray}{$\triangleright$\ }}{}
\newcommand{\mytcp}[1]{\tcp{\textcolor{gray}{#1}}}
\SetCommentSty{textnormal}
\SetKwProg{Fn}{Function}{}{}
\SetKw{Break}{\KwSty{break}}

\Input{Dataset $D$, task prompt \texttt{P}, an integer $k$}
\Output{Clustering of $D$}

\vspace{0.5em}

$m \gets 0$;  \texttt{should\_terminate}=\texttt{False}, $C^+, C^-\leftarrow0$\;

\While{\textbf{not} \texttt{should\_terminate} and $m < m_{\max}$}{
    $m \gets m+1$\;
    $W, C^+, C^-\leftarrow\texttt{UpdateEdgeWeights}(D, \texttt{P}, C^+, C^-, m)$\;
    $\hat{C}, \texttt{should\_terminate}\gets$ cluster $D$ using $W$, decide termination\;
}
\Return{$\hat{C}$}\;

\end{algorithm}
%

\vspace{-0.2cm}
\subsubsection{Computing Edge Weights.}\label{sec:clustering:edge_weights}
Here, we discuss how to compute the edge weights for record pairs $(s, t)\in E$. A straightforward approach is to use an LLM to estimate the probability that $s$ and $t$ belong to different classes, independently for each pair.
However, this strategy provides no dataset context to the LLM and yields poor estimates.

\textbf{Overview}. As mentioned earlier, to enable better holistic dataset understanding while providing high quality results, we use a bagging-like approach~\cite{breiman1996bagging}. We iteratively query an LLM whether the pair $(s, t)$ belongs to the same class, but to provide data context, at the $i$-th iteration, provide the LLM with an additional subset of the data $S_i\subseteq D$, sampled uniformly at random from $D$. We then average LLM answer across these calls as the edge weight. However, this approach is expensive since it requires $O(|D|^2)$ LLM calls across pairs, each costing $O(|S_i|)$ recoreds due to the additional data input for each call.  

To reduce cost, instead of querying the LLM for only $(s, t)$, we ask the LLM to output all pairs in $S_i$ that should belong to the same class, obtaining labels for multiple edges per LLM call. In this step, in addition to the pairs the LLM directly outputs, we also include pairs that are implied by transitivity based on LLM output (e.g., if LLM outputs $t_a$ and $t_b$, and $t_b$ and $t_c$  belong to the same class, we also assume $t_a$ and $t_c$ belong to the same class). We use the pairs in transitive closure of LLM output to compute edge weights.

\textbf{Algorithm}. At the $i$-th iteration, the LLM is given a set $S_i\subset D$ and asked to determine which pairs of records in $S_i$ belong to the same class. We let $P_i\subseteq S_i\times S_i$ the set of pairs output by the LLM and let $\bar{P}_i$ be its transitive closure. $\bar{P}_i$ is the set of \textit{positive annotations}, i.e., pairs that should belong to the same class. We denote by $\bar{N}_i=S_i\times S_i\setminus \bar{P}_i$ the rest of the pairs, i.e., the set of \textit{negative annotations} that shouldn't belong to the same class. After each iteration, we calculate the edge weights, $W$, as the frequency of negative annotations for each record pair. 

We calculate edge weights iteratively, after computing negative and positive annotations at the $i$-th iteration. Specifically, for $a, b\in [|D|]$, let $C_i^+[a, b]$ denote the number of positive annotations for $(t_a, t_b)\in E$ observed until the $i$-th iteration and $C_i^-[a, b]$ defined analogously for negative pairs. Formally, 
\vspace{-0.05cm}
\begin{align*}
C_i^+[a, b] &= C_{i-1}^+[a, b] + \mathds{I}[(t_a, t_b) \in \bar{P}_i], & C_0^+[a, b] &= 0, \\
C_i^-[a, b] &= C_{i-1}^-[a, b] + \mathds{I}[(t_a, t_b) \in \bar{N}_i], & C_0^-[a, b] &= 0.
\end{align*}
\vspace{-0.15cm}
Finally, $W_i[a, b] = \frac{C^-_i[a, b]}{C^+_i[a, b] + C^-_i[a, b]}$ is the edge weight at the $i$-th iteration. 
\if 
Let $C^+[a, b]=\sum_{i\in [m]} \mathds{I}[(t_a, t_b)\in \bar{P}_i]$ and $C^-[a, b]=\sum_{i\in [m]} \mathds{I}[(a, b)\in N_i]$ be the number of positive and negative labels across all samples $S_1, S_2, \cdots, S_m$, respectively.
We define the distance matrix $W$ by the frequency of negative labels, i.e.,
$W[a, b] = \frac{C^-[a, b]}{C^+[a, b] + C^-[a, b]}.$
\fi

\begin{figure}
\vspace{-0.7cm}
    \centering
    \includegraphics[width=0.86\columnwidth]{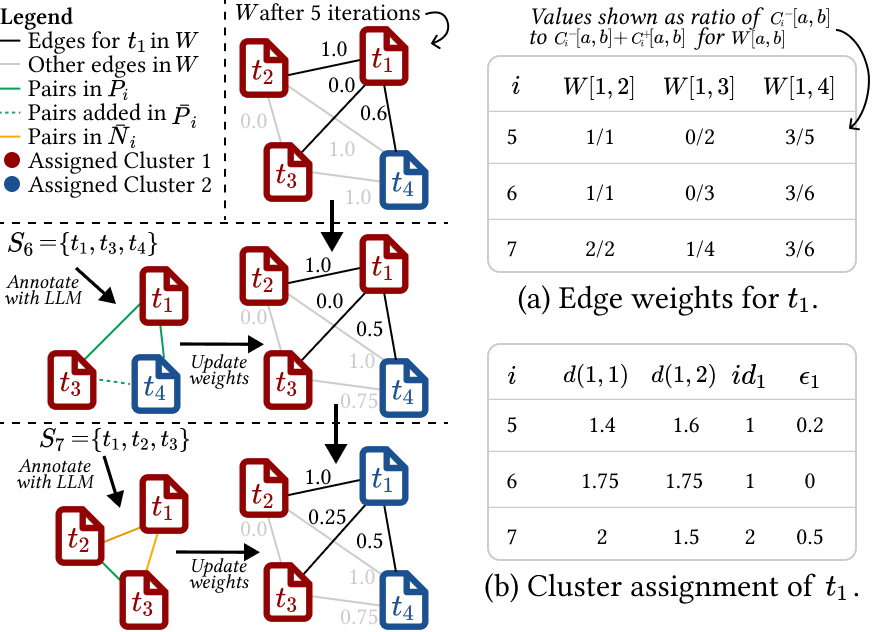}
    \caption{Cluster algorithm example.}
    \label{fig:pairwise}
\end{figure}
\begin{algorithm}[t]
\footnotesize
\caption{\texttt{UpdateEdgeWeights}$(D, \texttt{P}, C_{i-1}^+, C_{i-1}^-, i)$}
\label{alg:sampling}

\SetKwInOut{Input}{Input}
\SetKwInOut{Output}{Output}
\SetKwComment{tcp}{\textcolor{gray}{$\triangleright$}\ }{}
\newcommand{\mytcp}[1]{\tcp{\textcolor{gray}{#1}}}
\newcommand{\mytcpstar}[1]{\tcp*{\textcolor{gray}{#1}}}
\SetCommentSty{textnormal}
\SetKwProg{Fn}{Function}{}{}
    \Input{Dataset $D$, task prompt \texttt{P}, positive and negative counts, $C_{i-1}^+, C_{i-1}^-$ so far, iteration number, $i$}
    \Output{Updated pairwise edge weight matrix $W$}
    \vspace{0.5em}

    \mytcp{Sample and obtain LLM-suggested pairs}
    $S\gets$ random sample from $D$\;
    $P \gets \mathcal{M}(\texttt{P}, S)$ \mytcpstar{Pairs to be in the same class}
    $\bar{P} \gets $ Transitive closure of $P$\;
    $\bar{N} \gets S\times S\setminus \bar{P}$ \;
    
    \For{$t_a, t_b \in S, a \neq b$}{
        \mytcp{Aggregate positive and negative labels}
        $C_i^+[a, b] \gets C_{i-1}^+[a, b] + \mathds{I}[(t_a, t_b)\in \bar{P}]$\;
        $C_i^-[a, b] \gets C_{i-1}^-[a, b] + \mathds{I}[(t_a, t_b)\in \bar{N}]$\;
        \mytcp{Modify pairwise edge weights}
        $W_i[a, b] \gets \frac{C_i^-[a, b]}{C_{i}^-[a, b]+C_{i}^+[a, b]}$\;
    }
    \Return{$W_i$, $C_{i}^+$, $C_{i}^-$}

\end{algorithm}

Alg.~\ref{alg:sampling} illustrates this process. We construct a sample set $S$ and query the LLM (lines 1-2), take transitive closure and obtain negative labels (lines 3-4).
We then update the counts of positive and negative labels (lines 5–7)---note that $C^+$ and $C^-$ are initially zero (in Alg.~\ref{alg:semantic-cluster}).
Finally, we update the edge weights using the proportion of negative annotations (lines 8–9).

\textit{Example.} Fig.~\ref{fig:pairwise} illustrates how edge weights get updated. 
The first graph (Fig.~\ref{fig:pairwise} Top) shows edge weights after $5$ iterations and Table (a) shows how they are calculated for $t_1$, e.g.,
$W[1, 4] = \frac{3}{5} = 0.6$, At $i=5$, the records are clustered into two clusters, $\hat{C}_1 = \{t_1, t_2, t_3\}$ and $\hat{C}_2 = \{t_4\}$.
At the $6$-th iteration, the algorithm samples $S_6 = \{t_1, t_3, t_4\}$, and the LLM returns pairs $P_6 = \{(t_1, t_3), (t_1, t_4)\}$.
By performing transitive closure, all three record pairs in $S_6$ are considered positives.
Therefore, the augmented pairs are $\bar{P}_6 = P_6 \cup \{(t_3, t_4)\}$.
We then update the edge weights, e.g., $W[1, 4]$ is updated to $\frac{3}{6} = 0.5$. This process is repeated for another sample set $S_7$ at the 7-th iteration. 

\vspace{-0.1cm}
\subsubsection{Correlation Clustering and Termination.}\label{sec:clustering:corr}
This step performs clustering and decides whether more edge weight refinement is needed. We use correlation clustering~\cite{gionis2007clustering} to cluster the vertices given the edge weight matrix, $W$. 

\textbf{Correlation Clustering.} Correlation clustering assigns clusters to records to minimize the sum of \textit{disagreements} across all pairs. The disagreement between a pair of records $t_a,t_b\in D$ is defined as $W[a, b]$ if $t_a$ and $t_b$ are assigned to the same cluster and $(1 - W[a, b])$ otherwise.
Formally, correlation clustering assigns a cluster ID $id_a \in \{1, 2, \cdots, k\}$ to each record $t_a, a \in [|D|]$.
We define $d(a, j)$ as the sum of disagreements for record $t_a$ if $t_a$ is assigned to cluster $j$:
\vspace{-0.08cm}
$$
d(a, j) = \sum_{\overset{id_b = j,}{a \neq b}} W[a, b] + \sum_{\overset{id_b \neq j,}{a \neq b}} \left(1 - W[a, b]\right).\vspace{-0.15cm}
$$
The goal is to assign cluster IDs that minimize the sum of disagreement for all records:
$\sum_{a \in [|D|]} d(a, id_a)$.
Since correlation clustering is NP-Hard, we use local search as in ~\cite{gionis2007clustering}.
The algorithm starts with a random cluster assignment and greedily modifies assignments to improve the objective.



\textbf{Termination Decision}. We terminate our clustering procedure (i.e., loop in line 2 of Alg.~\ref{alg:semantic-cluster}) if a change in clustering for a large subset of the records is unlikely after more edge weight refinement. 

We let $d^*(a, i)=\mathds{E}[d(a, i)]$ for the $i$-th cluster and $a$-th record, $i\in [k]$, $a\in [|D|]$, with the expectation taken over our sampling process. Intuitively, Change in clustering is likely if at a specific iteration $d(a, i)$ differs significantly from $d^*(a, i)$. This is because $d(a, i)$ must eventually converge to $d^*(a, i)$, and if $|d(a, i)-d^*(a, i)|$ is large, this can significantly impact the objective. Thus, to decide whether we should terminate or not, we use the difference between $d(a, i)$ and $d^*(a, i)$ as a proxy---if the difference is sufficiently small, then the current clustering is unlikely to change in the future.

Formally, we say the $a$-th record, assigned to a cluster $id_a\in[k]$, is \textit{uncertain}  if $d^*(a, id_a)-d(a, id_a)\geq \epsilon_a$ and $d(a, i)-d^*(a, i)\geq \epsilon_a$ for all $i\in [k]\setminus \{id_a\}$, for an $\epsilon_a\geq 0$ specified later. This notion evaluates whether $d$ underestimate $d^*$ for the cluster $a$ is assigned to while overestimating $d^*$ for other clusters---a setting which likely implies clustering of $a$ will change with more observations. 
We use $\epsilon_a=\frac{1}{2}\min_{j\in[k]\setminus\{id_a\}}(d(a, id_a)-d(a, j))$ to help decide whether sampling error is \textit{low enough}; error more than $\epsilon_a$ likely causes a change in clustering because it changes the best cluster for record $a$. Using this notion of uncertainty, then, our goal is to take enough samples so that most records are no longer uncertain. The following lemma, based on simplifying statistical assumptions, provides an upper bound on the expected number of uncertain records.

\begin{lemma}\label{lemma:decidability}
    Given a clustering $\hat{C}_1, ..., \hat{C}_k$ of a dataset $D$, and assuming edge weights are obtained independently, the expected number of uncertain records after $r$ samples is at most $\sum_{a\in D}\exp{(2r\sum_{i\in [k]}\frac{\epsilon_a^2}{|\hat{C}_i|^2})}$.
\end{lemma}

Proof of Lemma~\ref{lemma:decidability} is based on an application of Hoeffding's inequality~\cite{hoeffding1994probability} and is deferred to \iftoggle{techreport}{our technical report~\cite{techrep}.}{Appx.~\ref{appx:proofs}.} The lemma makes an independence assumption which is not necessarily true. Nevertheless, we observed this statistical model to be sufficient to obtain a suitable termination decision in practice.  

Finally, to decide whether to terminate, we evaluate if the bound on the number of undecided records in Lemma~\ref{lemma:decidability} is below a threshold $\tau$, and if so, terminate. We set $\tau$ to a small portion of dataset size $|D|$ (set to $0.2\times |D|$ in our experiments), since when the fraction of uncertain pairs is small, the clustering outcome is unlikely to significantly change based on changes to edge weights.

\subsection{Assigning Clusters to Classes}

\begin{figure}
\vspace{-0.7cm}
    \centering
    \includegraphics[width=\columnwidth]{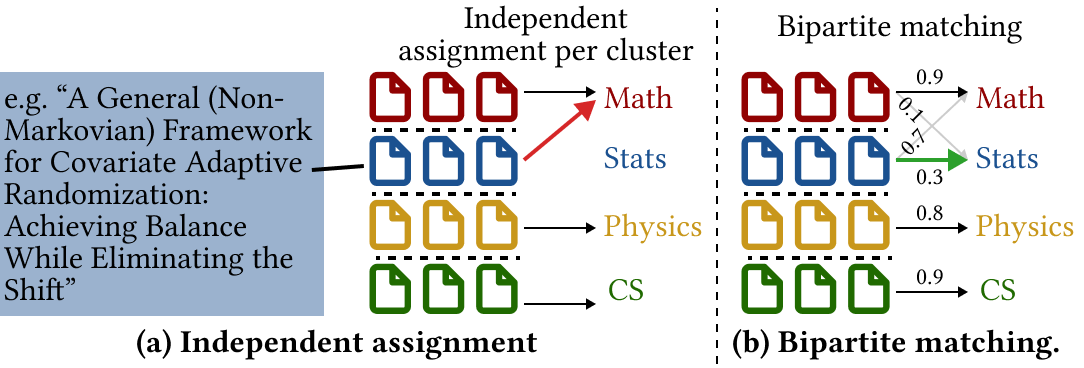}
    \caption{Cluster assignment example.}
    \vspace{0.1cm}
    \label{fig:transform}
\end{figure}

\label{sec:matching}
After obtaining clusters, we assign each to a class.
A simple strategy is to query the LLM for the most suitable class for each cluster independently.
However, this approach yields low accuracy especially when classes have overlapping boundaries. Errors occur due to a lack of holistic data understanding, now exacerbated because of potential inaccuracies in the clustering step. 
Fig.~\ref{alg:classification-match} shows an example task of classifying papers by topic. Classes include ``Math'' and ``Stats'' where ``Math'' class excludes statistics papers. Without seeing other math papers, an LLM may misclassify the statistics class as mathematics, as it cannot distinguish the boundary between them. A few misclassified records complicate the task, since a predominantly statistics cluster may contain a few non-statistics papers.

Instead of assigning class names independently, we take advantage of the fact that there is a one-to-one mapping between clusters and class names and jointly assign class names to all clusters using a bipartite matching approach.
Specifically, we construct a complete bipartite graph whose two partitions correspond to clusters and class names respectively, and the edge between a cluster and a class denotes the probability the cluster belongs to the class---we use LLMs to estimate this probability as discussed below. Then to assign clusters to class names, we find the maximum weight matching---maximizing the total probability of correct cluster assignment across all clusters---which can be done in PTime using the Hungarian Algorithm~\cite{kuhn1955hungarian}.


Formally, let $\hat{\mathcal{C}}=\{\hat{C_1}, ..., \hat{C_k}\}$ be the set of clusters, and consider the weighted bipartite graph $\mathcal{B} =(\hat{\mathcal{C}}, L, \hat{\mathcal{C}}\times L, W)$ with edge weights $W$. We use LLM log-probabilities when assigning class names to clusters as the edge weights $W$, specifically,
$$W(\hat{C}_i, \ell_j) = \mathscr{S}_{\mathcal{M}}(\texttt{P}, L, \hat{C}_i, \ell_j)\cdot |\hat{C_i}|,$$
where $\mathscr{S}_{\mathcal{M}}(\texttt{P}, L, \hat{C}_i, \ell_j)$ denotes the log-probability of model $\mathcal{M}$ when asked if cluster $\hat{C}_i$ belongs to class $\ell_j$. 
We then find a perfect matching $M\subseteq \hat{\mathcal{C}}\times L$ with maximum weights, $\sum_{(\hat{C}_i, \ell_j)\in M} W(\hat{C}_i, \ell_j)$. 

Algorithm ~\ref{alg:classification-match} shows this process. 
For each $\hat{C}_i$, we estimate the edge weights for all classes using an LLM $\mathcal{M}$ (lines 2-4)---this step can be implemented by a single LLM call for each cluster.
Then, we perform bipartite matching, obtain class names for each cluster (line 5) and then assign all the records in a cluster the class obtained for that cluster during bipartite matching (lines 6-8).

\begin{algorithm}[t]
\footnotesize
    \caption{$\texttt{Assign}(\hat{C},\texttt{P}, k, L)$}
    \label{alg:classification-match}

    \SetKwInOut{Input}{Input}
    \SetKwInOut{Output}{Output}
    \SetKwComment{tcp}{\textcolor{gray}{$\triangleright$\ }}{}
    \newcommand{\mytcp}[1]{\tcp{\textcolor{gray}{#1}}}
    \newcommand{\mytcpstar}[1]{\tcp*{\textcolor{gray}{#1}}}
    \SetCommentSty{textnormal}

    \Input{Clusters $\hat{C}_1, \hat{C}_2, \cdots, \hat{C}_k$, class labels $\{\ell_1, \ell_2, \cdots, \ell_k\}$, task prompt \texttt{P}}
    \Output{Class labels of all records}

    $W: \hat{C}\times \ell \to \mathbb{R}, W \gets0$\;

    \For{$i \gets 1$ \KwTo $k$}{
        \For{$\ell_j \in L$} {
            $W(\hat{C}_i, \ell_j) \gets \mathscr{S}_{\mathcal{M}}(\texttt{P}, L, \hat{C}_i, \ell_j) \cdot |\hat{C}_i|$\;
        }
    }
    $M \gets \texttt{BipartiteMatching}(\hat{C}, L, W)$

    \For{$(\hat{C}_i, \ell_{\hat{C}_i}) \in M$}{
        \For{$t_j \in \hat{C}_i$}{
            $\hat{y}_{j} \gets \ell_{\hat{C}_i}$\;
        }
    }

    \Return{ $\{\hat{y}_{1}, \hat{y}_{2}, \cdots, \hat{y}_{B}\}$ }
\end{algorithm}

\section{Budget-Aware Classification}
\label{sec:class-extend}
The clustering-based classification approach in Sec.~\ref{sec:class-method} can be expensive. As discussed in Sec.~\ref{sec:classification:overview}, we use a cascade approach to reduce the cost and adhere to budgetary constraints provided by the user (i.e., step \textcircled{2} of Algorithm~\ref{alg:classification-framework}). We first discuss our cascade approach in Sec.~\ref{sec:budget_aware:cascade}, and present a cost analysis in Sec.~\ref{sec:classification:analysis}.

\subsection{Cascade Approach}\label{sec:budget_aware:cascade}
In this section, we discuss our cascade approach to ensure the cost is within the budget $\widetilde{\mathcal{C}}$.

To maximize accuracy while staying within the cost budget $\widetilde{\mathcal{C}}$, for each record, we decide whether to use \texttt{CbClassification}, or row-by-row processing with $\mathcal{M}_e$ or $\mathcal{M}_c$. In model cascade terminology (see Sec.~\ref{sec:background}), we consider \texttt{CbClassification} as our oracle, and row-by-row processing with $\mathcal{M}_e$ or  $\mathcal{M}_c$ as two proxies, each with different cost-accuracy trade-offs. For a given dataset, we first decide a cascade architecture (i.e., what methods are used), and then decide how to set the cascade thresholds (which determines the records for which each method is used).

\textbf{Cascade Architecture}.
For simplicity, for a given dataset, we complement \texttt{CbClassification} (oracle) only with one of $\mathcal{M}_e$ or $\mathcal{M}_c$, that is, we pick only one of the proxies for our cascade. Thus, our cascade architecture processes some of the records with one of $\mathcal{M}_e$ and $\mathcal{M}_c$, and uses \texttt{CbClassification} for the rest. We use a worst-case estimate to decide which proxy model to use: we use $\mathcal{M}_e$ if there exists a cascade solution using only $\mathcal{M}_e$ and \texttt{CbClassification} that respects the cost budget, and use $\mathcal{M}_c$ otherwise. Formally, we use $\mathcal{M}_e$ if $\mathcal{C}_0 + \sum_{t\in \mathcal{D} \setminus D_0}\texttt{cost}(\mathcal{M}_e(\texttt{P}, t))\leq \widetilde{\mathcal{C}}$ and $\mathcal{M}_c$ otherwise. This decision criteria guarantees that, in the worst case,  we can process all records with the chosen model (since we can always set an appropriate threshold), though, in practice, we improve accuracy using \texttt{CbClassification}. More complex cascade architectures (e.g., using both models combined with \texttt{CbClassification}, each for different subsets) are possible but we observed our simple heuristic to provide good cost-accuracy trade-offs. 

\textbf{Cascade Threshold.} Denote by $\mathcal{M}^P\in\{\mathcal{M}_c, \mathcal{M}_e\}$ the chosen model based on the decision criteria discussed above. To process the records, we use $\mathcal{M}^P$'s confidence score for each record $t\in D$, denoted as $\mathscr{S}_{\mathcal{M}^P}(\texttt{P}, t)$, to decide whether to use the proxy for that record, similar to ~\cite{patel2024lotus, zeighami2025cut}. 
Specifically, we determine a threshold, $\tau^*$ on $\mathcal{M}^P$'s confidence scores, use \texttt{CbClassification} on records where $\mathcal{M}^P$'s confidence is below $\tau^*$ and use $\mathcal{M}^P$ on the rest. 
We pick a threshold that maximizes use of \texttt{CbClassification} while respecting the cost budget. For any candidate threshold, $\tau$, we define $\texttt{cost}(\tau)$ to be the total cost if we use row-by-row classification by $\mathcal{M}^P$ on all records with confidence at least $\tau$, and use \texttt{CbClassification} for the remaining.
We choose the largest $\tau$ such that $\texttt{cost}(\tau) \leq \widetilde{\mathcal{C}}$. 

It remains to estimate the cost, $\texttt{cost}(\tau)$, which consists of two parts: the cost of (1)  \texttt{CbClassification} on $D_0$ and records in $\mathcal{D} \setminus D_0$ with confidence less than $\tau$; and (2) row-by-row classification by $\mathcal{M}^P$ on $\mathcal{D} \setminus D_0$.
For (1) we use the cost on the sample batch $D_0$ to estimate the cost of \texttt{CbClassification} per batch as $\mathcal{C}_0$ (see Alg.~\ref{alg:classification-framework}). For (2) we compute the cost, $\mathcal{C}_{\mathcal{M}^P}$, of using $\mathcal{M}^P$ on $\mathcal{D} \setminus D_0$ 
based on number of tokens that would be passed to the LLM. $\texttt{cost}(\tau)$ is
\begin{align}\label{eq:cost}
\texttt{cost}(\tau) = \mathcal{C}_{\mathcal{M}^P} + \mathcal{C}_0 \cdot \left(1 + \lceil\frac{n(\tau)}{B}\rceil\right),
\end{align}
where $n(\tau) = |\{t \in \mathcal{D} \setminus D_0 \mid \mathscr{S}_\mathcal{M}(\texttt{P}, t) < \tau\}|$ denotes the number of records processed by \texttt{CbClassification} in $\mathcal{D} \setminus D_0$

\textbf{Algorithm}. Algorithm~\ref{alg:cascading} shows this process. We first check whether we can perform \texttt{CbClassification} on the full dataset (line 1).
If not, we estimate the cost $\mathcal{C}_{\mathcal{M}_e}$ (line 3).
If $\mathcal{M}_e$ is feasible, we choose it as the proxy (line 4).
Otherwise, we choose $\mathcal{M}_c$ as the proxy (line 5).
Then, we calculate the threshold $\tau$ based on the chosen proxy and the cost-estimation formula in Eq.~\ref{eq:cost} (line 6).
Finally, we filter out the records with confidence scores above $\tau$ as $D_R$ (line 7), and return row-by-row assignment from $\mathcal{M}^P$ as their predictions (line 8).

\begin{algorithm}[t]
\footnotesize
    \caption{$\texttt{PredictWithCascade}(D, \texttt{P}, k, L, D_0, \mathcal{C}_0)$}
    \label{alg:cascading}

    \SetKwInOut{Input}{Input}
    \SetKwInOut{Output}{Output}
    \SetKwComment{tcp}{\textcolor{gray}{$\triangleright$\ }}{}
    \newcommand{\mytcp}[1]{\tcp{\textcolor{gray}{#1}}}
    \newcommand{\mytcpstar}[1]{\tcp*{\textcolor{gray}{#1}}}
    \SetCommentSty{textnormal}

    \Input{Dataset $D$, task prompt \texttt{P}, integer $k$, class labels $L$, sample batch $D_0$, cost $\mathcal{C}_0$}
    \Output{Classification $\hat{y}_{D_R}$ of the subset $D_R\subseteq D\setminus D_0$}

    \lIf{$\mathcal{C}_0 \cdot \lceil\frac{n}{B}\rceil\leq \widetilde{\mathcal{C}}$}{\Return{$\varnothing$}}
    \Else{
        $\mathcal{C}_{\mathcal{M}_e} \gets \sum_{t\in \mathcal{D} \setminus D_0}\texttt{cost}(\mathcal{M}_e(\texttt{P}, t))$\;
    
        \lIf{$\mathcal{C}_0 + \mathcal{C}_{\mathcal{M}_e} \leq \widetilde{\mathcal{C}}$}{
            $\mathcal{M}^P\gets \mathcal{M}_e$
        }
        \lElse{
            $\mathcal{M}^P\gets \mathcal{M}_c$
        }
        $\tau \gets \max_{0 \leq \tau \leq 1} \{\tau \mid \texttt{cost}(\tau) \leq \widetilde{\mathcal{C}}\}$\;
        $D_R \gets \{t \mid t \in \mathcal{D} \setminus D_0, \mathscr{S}_{\mathcal{M}}(\texttt{P}, t) \geq \tau\}$\;
        \Return {$\{\mathcal{M}^P(\texttt{P}, t, L) \mid t \in D_R\}, D_R$}
    }
  




\end{algorithm}

\subsection{Cost Analysis and Discussion}\label{sec:classification:analysis}

\textbf{Cost Analysis.} We analyze the cost of \sys for classification. To simplify notation, we assume all records and class names have fixed length, and denote by $L_r$ total number of tokens in the dataset and by $L_\ell$ the total number of tokens in the class names. Also denote by $m$ the number of sampling iterations in Alg.~\ref{alg:semantic-cluster} (i.e., number of iterations of the loop in line 2), $r=\frac{|D_X|}{n}$ the fraction of records processed by clustering-based classification, and $c_{\text{proxy}}$ as the cost per token for the model $\mathcal{M}^{P}$ chosen as proxy in Sec.~\ref{sec:budget_aware:cascade}. All of $m, r$ and $c_{\text{proxy}}$ are automatically determined by \sys. Finally, let $c_{\textnormal{assignment}}$ and $c_{\textnormal{cluster}}$ denote the cost per token for models used in our clustering and cluster assignment steps, whose choice is determined at the end of this section based on this cost analysis. The following proposition shows the cost of \sys.

    \vspace{-0.1cm}
\begin{proposition}\label{prop:cost}
The total cost of classification with \sys (i.e., Alg.~\ref{alg:classification-framework}) is at most
    \vspace{-0.06cm}
$$\varkappa L_r(c_{\textnormal{proxy}}+mrc_{\textnormal{cluster}}+ rc_{\textnormal{assignment}}) + \varkappa nL_\ell( c_{\textnormal{proxy}}+rkc_{\textnormal{assignment}}), $$\vspace{-0.1cm}
where $\varkappa$ is a universal constant.
\end{proposition}
The proof is deferred to \iftoggle{techreport}{our technical report~\cite{techrep}.}{Appx.~\ref{appx:proofs}.}

\textbf{Discussion.} First, in the above expression, the first term involving $L_r$ is usually dominant, since the number of tokens per record is typically much larger than the number of tokens in class names. Second, the cost depends on $m$, that is the number of samples taken during clustering. In practice, $m$ depends on the difficulty of clustering for a given dataset. If clustering is simple for a dataset, often few samples to obtain edge weights are sufficient to obtain well-separated clusters, so \sys terminates early (based on the decision criteria in Sec.~\ref{sec:clustering:corr}). On the other hand, for a difficult dataset, clusters may not be well-separated, so minor changes in edge weights can have significant impact on clustering, and thus \sys terminates after taking more samples. As such, \sys adapts to the dataset, incurring more cost if dataset is difficult. Finally, we use $\mathcal{M}_c$ for clustering (in Alg.~\ref{alg:sampling}) and $\mathcal{M}_e$ for cluster assignment (in Alg.~\ref{alg:classification-match}) to ensure cost remains low even for large $m$.


\if 0\textbf{Parameter Setting}.
In practice, record texts are typically much longer than ground-truth class names.
Moreover, to achieve high clustering accuracy on each batch $D_0$ and $D_{X, i}$ with $B = \max\{200, 10\times k\}$, the required number of samples is approximately $m = O(B)$.
As a result, the sampling cost dominates the overall cost.
To balance cost and accuracy, we use  $\mathcal{M}_1 = \mathcal{M}_c$ for clustering, $\mathcal{M}_2 = \mathcal{M}_e$ for cluster assignment.
The choice of $\mathcal{M}$ to be either $\mathcal{M}_e$ or $\mathcal{M}_c$ is decided in Sec.~\ref{sec:budget_aware:cascade}.
\fi

\section{Other Tasks with \sys}
\label{sec:others}

\begin{algorithm}[t]
\footnotesize
    \caption{$\texttt{\sys}_{\texttt{T}}(\mathcal{D}, \texttt{P}, k, L)$}
    \label{alg:all-op-framework}

    \SetKwInOut{Input}{Input}
    \SetKwInOut{Output}{Output}
    \SetKwComment{tcp}{\textcolor{gray}{$\triangleright$\ }}{}
    \newcommand{\mytcp}[1]{\tcp{\textcolor{gray}{#1}}}
    \SetCommentSty{textnormal}
    \SetKwSwitch{Switch}{Case}{Other}{switch}{do}{case}{otherwise}{end}

    \Input{Task type \texttt{T}, Dataset $\mathcal{D}$, task prompt \texttt{P}, integer $k$, class labels $L$ ($\varnothing$ when clustering)}
    \Output{Answer $\hat{Y}$}

  
    \mytcp{\textcircled{1} Cluster and assign a sample batch $D_0$}
    $D_0 \gets$ a sample batch from $\mathcal{D}$\;
    $\hat{C} \gets \texttt{Cluster}(D_0, \texttt{P}, k)$\;

    \lIf{$\texttt{T}=\text{clustering}$}{$L \gets$ summarize clusters with LLM}

    \Switch{\texttt{T}}{
        \Case{$\text{classification or clustering}$}{
            $\hat{Y}_{D_0} \gets \texttt{Assign}(\hat{C}, \texttt{P}, k, L)$
        }
        \lCase{$\text{scoring}$}{
            $\hat{Y}_{D_0} \gets \texttt{Sort}(\hat{C}, \texttt{P}, k, L)$
        }
    }

    $\mathcal{C}_0 \gets$ total cost of step \textcircled{1}\;

    \mytcp{\textcircled{2} Model cascade}

    $\hat{Y}_{D_R} \gets \texttt{PredictWithCascade}(\mathcal{D}, \texttt{P}, k, L, D_0, \mathcal{C}_0)$
    



    
    \mytcp{\textcircled{3} Batched processing}

    $\hat{Y}\gets \hat{Y}_{D_0} \cup \hat{Y}_{D_R}$\;
    $D_{X, 1}, \cdots, D_{X, n_B} \gets$ batches partitioned from $ \mathcal{D}\setminus D_0 \setminus D_R$\;

    \For{$i \gets 1$ \KwTo $n_B$}{
        $\hat{C}_{X, i} \gets \texttt{Cluster}(D_{X, i}, \texttt{P}, k, L)$\;
        \Switch{\texttt{T}}{
            \Case{$\text{classification or clustering}$}{
                $\hat{Y} \gets \hat{Y}\cup \texttt{Assign}(\hat{C}_{X, i}, \texttt{P}, k, L)$
            }
            \lCase{$\text{scoring}$}{
                $\hat{Y} \gets \hat{Y}\cup \texttt{Sort}(\hat{C}_{X, i}, \texttt{P}, k, L)$
            }
        }
    }

    \Return{$\hat{Y}$}
\end{algorithm}
In this section, we discuss how \sys can be used to perform other tasks. We first discuss scoring and clustering tasks in Sec.~\ref{sec:other-scoring}, for which a generalization of our classification method can be used. We then discuss other tasks such as extraction, joins and summarization in  Sec.~\ref{sec:other-tasks}.

\vspace{-0.13cm}
\subsection{Scoring and Clustering with \sys}
\label{sec:other-scoring}
We discuss extending our solution to scoring and clustering. We first define scoring and clustering problems in Sec.~\ref{sec:scoring_clustering:def}, and present a unified algorithm that solves all of classification, scoring, and clustering tasks in Sec.~\ref{sec:other_tasks:unified}, with task-specific details in Sec.~\ref{sec:other_tasks:scoring}.

\vspace{-0.13cm}
\subsubsection{Problem Definition\\\nopunct}\label{sec:scoring_clustering:def}
\textbf{Scoring}. Scoring is similar to classification, except that we assign an integer score to every record, so that \textit{classes are ordered}. Specifically, the user provides a natural language prompt \texttt{P} specifying the scoring instruction, an integer $k$ so that the set of possible scores is $\{1, 2, ..., k\}$, and an (optional) description of the $k$ scores $\ell_1, \ell_2, \cdots, \ell_k$. 
The goal is to assign a score, $\hat{z}_i\in[k]$ to each record $i \in [n]$. 

\textbf{Clustering}. 
The user provides a natural language prompt \texttt{P} specifying the clustering task, and an integer $k$ denoting the number of clusters.
The goal of clustering is to partition $\mathcal{D}$ into $k$ disjoint predicted clusters $\{\hat{C}_1, \hat{C}_2, \cdots, \hat{C}_k\}$ according to \texttt{P}.

\vspace{-0.13cm}
\subsubsection{Unified Algorithm for Classification, Scoring, and Clustering}\label{sec:other_tasks:unified}
We make minor modifications to our classification algorithm to perform scoring and clustering. As was illustrated in Fig.~\ref{fig:classification-framework}, our classification algorithm \textcircled{1} classifies a sample subset of the data using a clustering-based classification method, \textcircled{2} uses a cascade approach to decide what subset of the data to classify with cheaper row-by-row methods and \textcircled{3} processes the rest of the data with the clustering-based classification method. 

Scoring is done almost identically to classification, but we take advantage of the order among the scores to improve cluster assignment in steps \textcircled{1} and \textcircled{3}. Specifically, instead of using bipartite matching, we use an ILP-based sorting method to assign clusters to scores. We use an LLM to decide the order for every pair of clusters, establish a total order across all clusters that minimizes violations of the pairwise orders obtained from the LLM, and assign each cluster a score according to its position in the order (details in Sec.~\ref{sec:other_tasks:scoring}). Readers familiar with work on crowd-powered algorithms will notice similarities to these approaches~\cite{marcus2011human, guo2012so, conitzer2006improved}.

Clustering also follows our classification algorithm, except after we cluster the initial batch of records,  $D_0$, we use an LLM to summarize each cluster. We then consider these cluster summaries as class names in a classification task, and classify the rest of the data into the classes specified by the summaries. The clusters are finally the records classified to belong to the same cluster summary. 

Algorithm~\ref{alg:all-op-framework} describes the workflow for performing classification, scoring, and clustering under this framework.
In step \textcircled{1}, we cluster a sample batch (lines 1-2) and assign clusters to labels with a task dependent assignment algorithm (lines 3-7).
For clustering, this step also generates labels (i.e., cluster summaries)  (line 3).
We then record the total cost of step \textcircled{1} (line 8) and use cascade for row-by-row LLM assignment (line 9).
Finally, we split the remaining records into batches (line 11). Each batch is first clustered (line 13), and then assigned according to the task (lines 14-17).

\subsubsection{Assigning Scores to clusters}\label{sec:other_tasks:scoring}
For scoring, our bipartite cluster assignment algorithm presented in Sec.~\ref{sec:matching} performs poorly because LLMs are unable to accurately estimate the probability of a cluster belonging to a specific score. Thus, the edge weights in the bipartite graph would be inaccurate, leading to incorrect assignment. Instead, we take advantage of the fact that clusters must be ordered to perform cluster assignment. Rather than asking directly for scores from LLMs, we instead ask for comparison between clusters, i.e., whether a cluster should be scored higher or lower than another---similar to prior work~\cite{liu2024aligning}, we observed LLMs to provide better pairwise comparison between clusters than directly scoring each. We then use these pairwise comparisons between clusters to establish a total order between them, where we find an order with minimum violations of the pairwise comparisons. 

Formally, for every pair of clusters, $\hat{C}_i$ and $\hat{C}_j$, we first use an LLM to estimate the probability that $\hat{C}_i$ has a higher score than $\hat{C}_j$, denoted by $W(\hat{C}_i, \hat{C}_j)$. To obtain $W(\hat{C}_i, \hat{C}_j)$, we randomly sample record pairs $(s,t)\in \hat{C}_i\times\hat{C}_j$, ask an LLM to decide which record, $s$ or $t$ should be scored higher,  and let $W(\hat{C}_i, \hat{C}_j)$ to be the fraction of pairs where $s$ was judged to have a lower score than $t$ by the LLM. 

We then find a total ordering of clusters such that the total weight of incorrect pairwise orders is minimized. That is, if $\pi(\hat{C}_i)$ denotes the score assigned to $\hat{C}_i$, we minimize
$$
\sum_{1 \leq i < j \leq k} \mathds{I}[\pi(\hat{C}_i) > \pi(\hat{C}_j)] W(\hat{C}_i, \hat{C}_j) + \mathds{I}[\pi(\hat{C}_i) < \pi(\hat{C}_j)]  W(\hat{C}_j, \hat{C}_i).
$$
We formulate the problem of finding a total ordering while minimizing the above objective as an Integer Linear Program (ILP) and solve it optimally, details are deferred to \iftoggle{techreport}{our technical report~\cite{techrep}.}{Appx.~\ref{appx:ilp}.}

\begin{algorithm}[t]
\footnotesize
    \caption{$\texttt{Sort}(\hat{C},  \texttt{P}, k, L)$}
    \label{alg:scoring-sort}

    \SetKwInOut{Input}{Input}
    \SetKwInOut{Output}{Output}
    \SetKwComment{tcp}{\textcolor{gray}{$\triangleright$}\ }{}
    \newcommand{\mytcp}[1]{\tcp{\textcolor{gray}{#1}}}
    \newcommand{\mytcpstar}[1]{\tcp*{\textcolor{gray}{#1}}}
    \SetCommentSty{textnormal}

    \Input{Clusters $\hat{C}_1, \hat{C}_2, \cdots, \hat{C}_k$, task prompt \texttt{P}, integer $k$ score labels $L$}
    \Output{Scores for records in $D$}
     $W\leftarrow 0$\;
    \For{$i, j \gets 1$ \KwTo $k$}{
            \For{$it \gets 1$ \KwTo $m_{\text{sort}}$}{
                $s, t \gets $ random record pair from $\hat{C}_i\times \hat{C}_j$\;
                $order \gets \mathcal{M}(\texttt{P}, s, t)$ \mytcpstar{Decide order between $s$ and $t$}
                \lIf{$order = \texttt{`<'}$}{$W(\hat{C}_i, \hat{C}_j) \gets W(\hat{C}_i, \hat{C}_j) + 1$}
            }
            $W(\hat{C}_i, \hat{C}_j)\gets \frac{1}{m_{sort}}W(\hat{C}_i, \hat{C}_j)$\;
    }

    $\pi \gets \texttt{ILP}(W)$\mytcpstar{Find total order across clusters}
    
    \For{$i \gets 1$ \KwTo $k$}{
        \For{$t_j \in \hat{C}_i$}{
            $\hat{z}_{j} \gets \pi(\hat{C}_i)$\;
        }
    }

    \Return{ $\{\hat{z}_{1}, \hat{z}_{2}, \cdots, \hat{z}_{B}\}$ }
\end{algorithm}

Algorithm~\ref{alg:scoring-sort} shows the cluster assignment process. We iterate over each pair of clusters to obtain likelihood estimate of their pairwise order  (lines 1-6).
We then use an ILP to obtain the scoring $\pi$ that minimizes the sum of violations (line 7).
Finally, we assign all records in the cluster to the same score based on $\pi$ (lines 8-10).

\vspace{-0.1cm}
\subsection{Other Tasks}
We next briefly discuss how \sys can be used for other data processing tasks. First, note that, as described, \sys directly applies to filtering tasks---a filter is a binary classification task. We expect similar ideas to also apply to joins, since a join is a  filter on the cross product of two datasets. However, we expect our cascade approach needs to change to allow for better cost reduction, following existing cost optimizations used for joins~\cite{patel2024lotus, zeighami2025featurized}. 

Finally, we expect other open-ended map operations, e.g., extraction or summarization tasks, to also benefit from holistic data understanding. For example, for extraction, holistic data understanding can help resolve ambiguities introduced by formatting or text alignment issues (e.g., by understanding correct alignment from other documents), or it can help decide correct output format and content (e.g., when extracting address, whether the country should be included or not). Similar benefits can also hold for summarization. For example, when first clustering records and asking an LLM to summarize each, a suitable summarization may need to compare and contrast each cluster which requires holistically understanding the data. A more thorough exploration is out of scope of this paper and is left to future work. 

\label{sec:other-tasks}

\vspace{-0.1cm}
\section{Experimental Results}
\label{sec:experiments}

We next empirically evaluate \sys. 
We discuss the experiment settings in Sec.~\ref{sec:experiments:setup}, compare \sys with baselines in Sec~\ref{sec:experiments:main}, and present ablation studies in Sec.~\ref{sec:expreriments:ablation}. Additional ablation studies and results on sensitivity to system parameters are presented in \iftoggle{techreport}{our technical report~\cite{techrep}.}{Appx.~\ref{appx:exp}.} 

\vspace{-0.1cm}
\subsection{Setup}
\label{sec:experiments:setup}

\textbf{Tasks and datasets.}
We conduct experiments on a diverse collection of 15 real-world datasets typically used for classification, scoring, and clustering. For classification and clustering, datasets used are listed in Table~\ref{tab:datasets_classification}.
These datasets cover a wide range of domains, including news, Reddit comments, and academic papers.
Each dataset contains two columns: the text data and a label. For classification, the goal is to predict this label. For clustering, following standard practice~\cite{xie2016unsupervised,guo2017improved}, we use the same datasets but withhold the label set from all methods. We then evaluate how well each method's clustering agrees with the clustering induced by the labels.  
In both cases, task prompt is a short sentence, e.g., ``Classify news articles based on their topic.'' For the scoring task, we use datasets listed in Table~\ref{tab:datasets_scoring}. These datasets span topics including essay scoring, sentiment analysis, and product analysis.
We provide the text data, the scoring range, and the description of each score as labels to the methods. Prompts for all datasets and further details are provided in \iftoggle{techreport}{our technical report~\cite{techrep}.}{Appx.~\ref{appx:exp}.} 

\begin{figure}[t]
\vspace{-0.8cm}
\centering
\captionof{table}{Summary of dataset statistics}
\label{fig:real_datasets}

\begin{subfigure}[t]{\columnwidth}
\centering
\footnotesize
\subcaption{Datasets for classification and clustering}
\label{tab:datasets_classification}
\vspace{-0.2cm}

\setlength{\tabcolsep}{3pt}
\begin{tabular}{l c c c c}

\toprule
\textbf{Dataset} & \textbf{Size} & $\mathbf{k}$ & \textbf{Description} & \textbf{Classes} \\
\midrule

AgNews \cite{zhang2015character}   & 127,600 & 4  & News stories & Section \\
Blurb~\cite{aly2019hierarchical}       & 28,729  & 36  & Book blurbs & Generes \\
Clinc \cite{larson2019evaluation}   & 22,500  & 10 & User queries & Domain \\
DBPedia \cite{zhang2015character}  & 630,000 & 14 & Wiki entries & Category \\
GoEmo \cite{dorottya2005goemotions} & 211,225 & 27 & Reddit comments & Emotion \\
Massive \cite{muennighoff2023mtebmassivetextembedding} & 22,288 & 11 & User queries & Domain \\
Mtop \cite{muennighoff2023mtebmassivetextembedding}    & 16,521 & 18 & User queries & Domain \\
ArxivS2S \cite{muennighoff2023mtebmassivetextembedding}   & 732,723 & 38 
& Paper titles & Primary domain \\
BiorxivS2S \cite{muennighoff2023mtebmassivetextembedding} & 75,000  & 25 & Paper titles & Primary domain\\
MedrxivS2S \cite{muennighoff2023mtebmassivetextembedding} & 37,500  & 51 & Paper titles & Primary domain\\

\bottomrule
\end{tabular}
\end{subfigure}
\hfill
\begin{subfigure}[t]{\columnwidth}
\centering
\subcaption{Datasets for scoring.}
\label{tab:datasets_scoring}
\vspace{-0.2cm}

\footnotesize
\setlength{\tabcolsep}{2pt}

\begin{tabular}{l c c c c}
\toprule
\textbf{Dataset} & \textbf{Size} & $\mathbf{k}$ & \textbf{Description} & \textbf{Scoring dimension}\\
\midrule

Amazon \cite{muennighoff2023mtebmassivetextembedding}       & 701,528 & 6 & Product reviews & Rating\\
Asap \cite{hamner2012asap_aes}          & 1,982 & 4 & Student essays & Writing grade\\
Sentiment \cite{eladjelet_abdelmalek_2025}     & 214,158 & 3 & User comments & Sentiment polarity\\
TicketSupport \cite{suraj520_customer_support_ticket} & 8,469 & 4 & Support tickets & Priority\\
Yelp \cite{muennighoff2023mtebmassivetextembedding}         & 50,000 & 5 & Business reviews & Rating\\

\bottomrule
\end{tabular}
\end{subfigure}

\end{figure}

\begin{figure*}
\vspace{-0.8cm}
    \hspace{-0.5cm}
    \begin{minipage}{0.62\textwidth}
        \includegraphics[width=\linewidth]{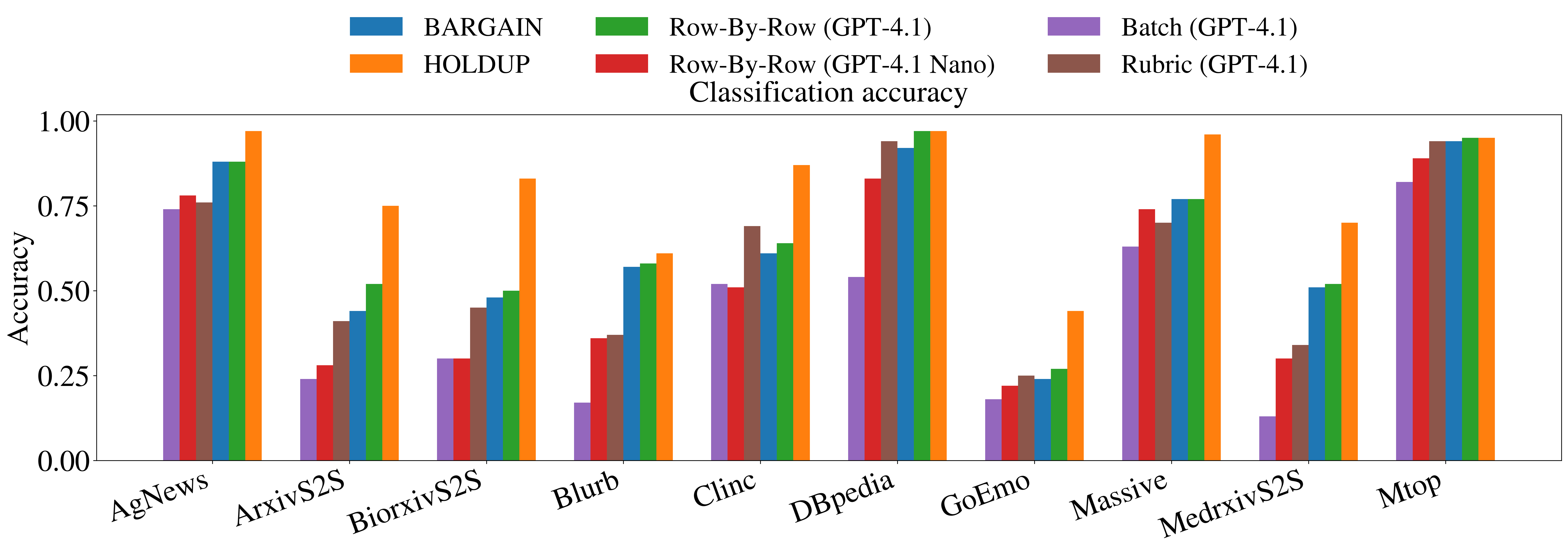}
        \caption{Classification accuracy across datasets}
        \label{fig:class-acc}
    \end{minipage}
    \hfill
    \begin{minipage}{0.40\textwidth}
        \includegraphics[width=1.05\linewidth]{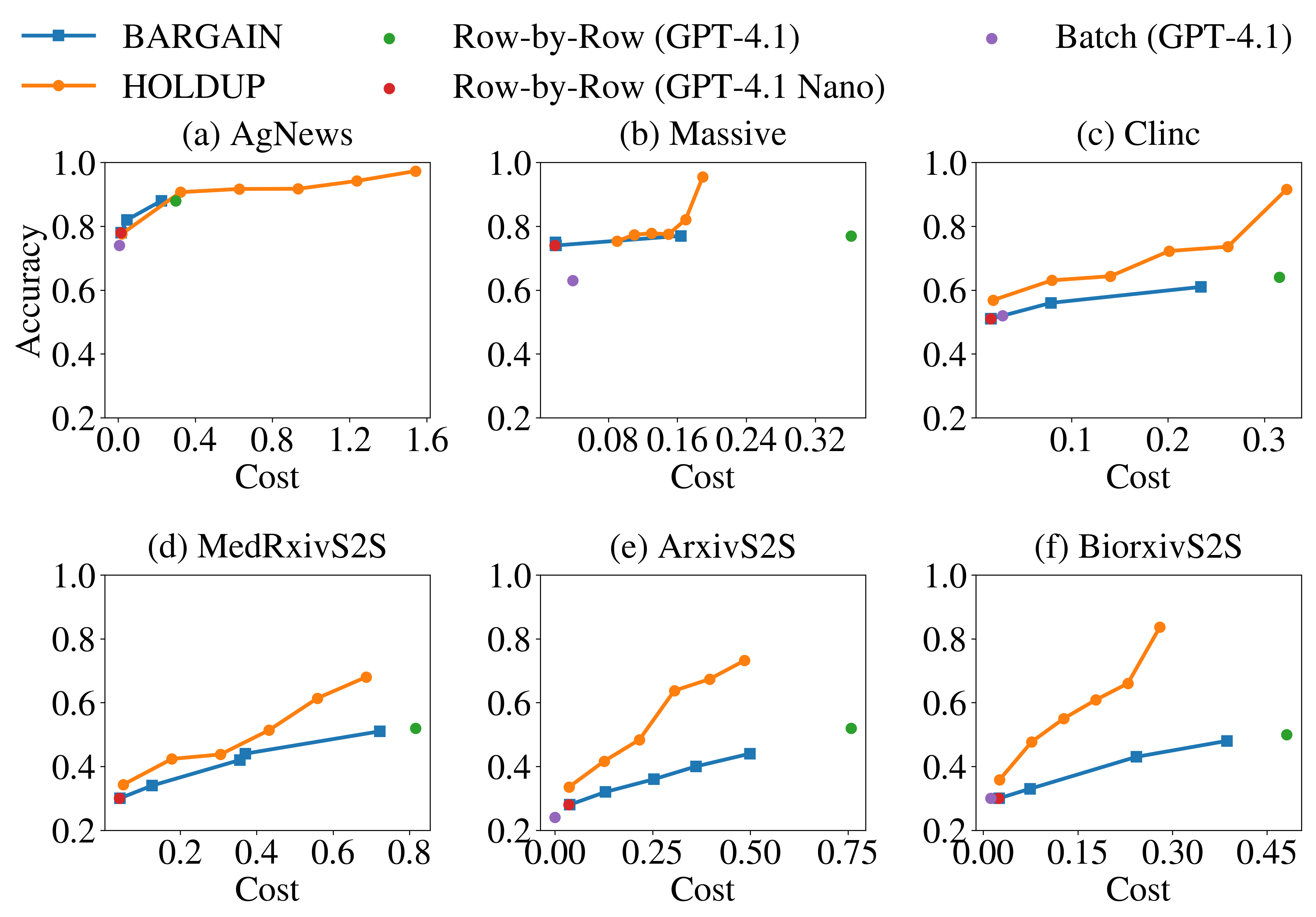}
        \caption{Cost/accuracy trade-offs for classification}
        \label{fig:cost}
    \end{minipage}
\end{figure*}

\begin{figure*}
    \hspace{-0.5cm}
    \begin{minipage}{0.61\textwidth}
    \includegraphics[width=\textwidth]{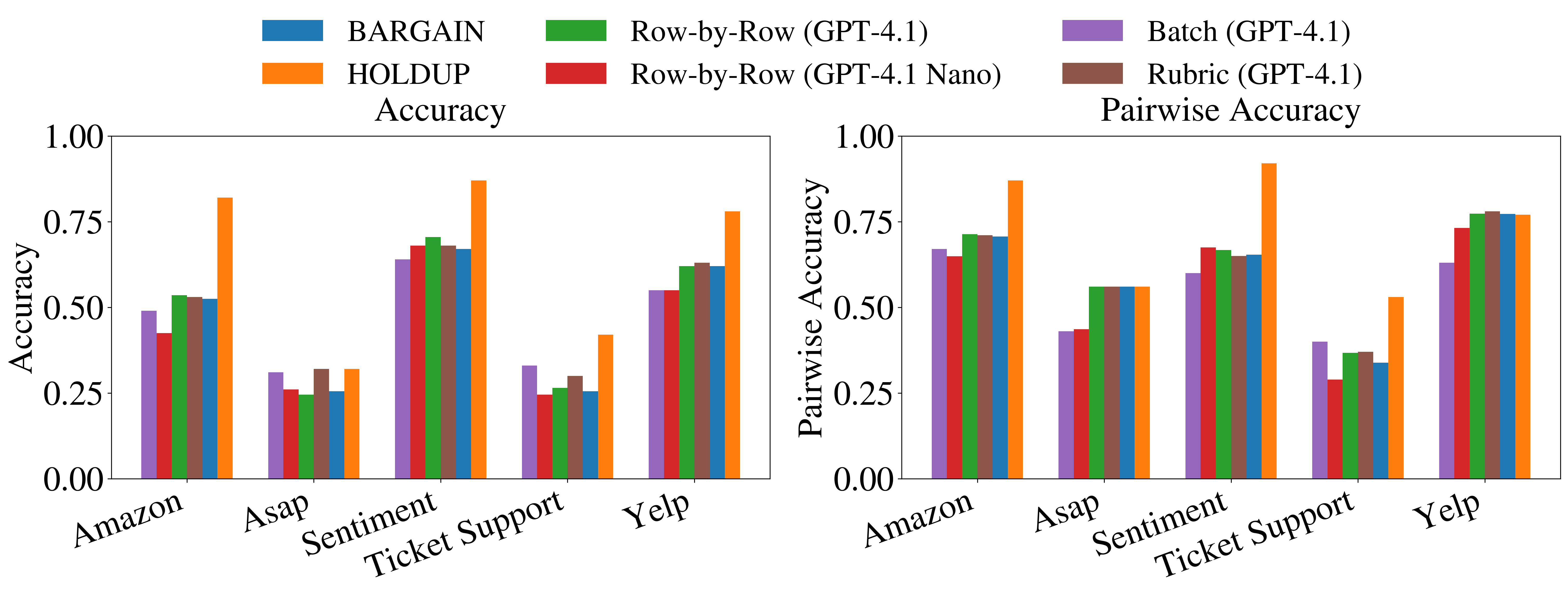}
    \caption{Scoring accuracy across datasets}
    \label{fig:scoring-acc}
    \end{minipage}
    \hfill
    \begin{minipage}{0.41\textwidth}
        \includegraphics[width=1\linewidth]{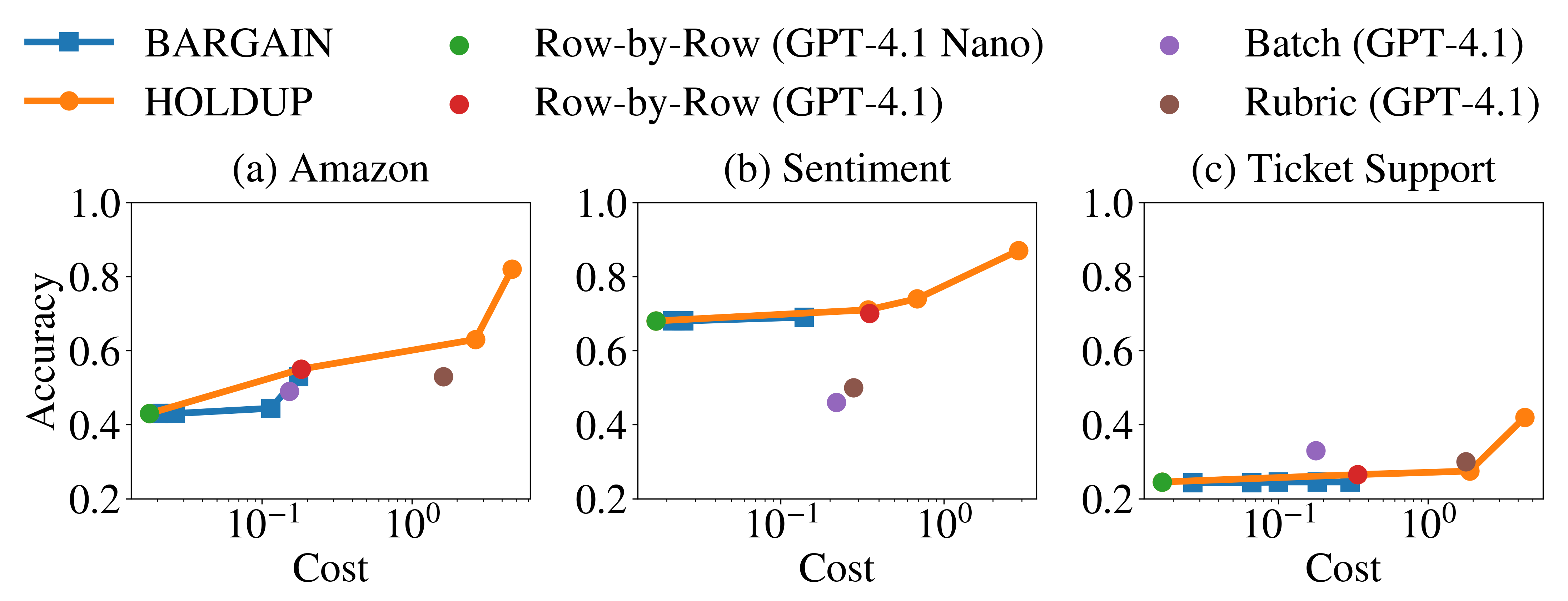}
        \caption{Cost/accuracy trade-offs for scoring}
        \label{fig:scoring-cost}
    \end{minipage}
\end{figure*}

\textbf{Baselines.} For classification and scoring, we compare \sys against three classes of baselines. The first involves \textit{Row-by-Row} processing, which follows existing systems~\cite{patel2024lotus, shankar2024docetl, russo2025abacus, jo2024thalamusdb} that apply an LLM independently to each row given the task instruction. The second class considers simple methods for incorporating data context during processing. \textit{Batch} partitions the dataset into batches and applies an LLM independently to each one, while \textit{Rubric} first provides a random sample of records to an LLM to generate a task rubric, then applies the LLM row by row with this rubric appended to the prompt. The third class considers methods with cost optimization. We use \textit{BARGAIN}~\cite{zeighami2025cut}, the state-of-the-art model cascade method that optimizes cost while still processing rows independently, to evaluate cost-accuracy tradeoffs.

For clustering, we compare against various baselines that use LLMs or embedding models for clustering: (1) \textit{K-means+Embed} applies K-means to record embeddings, a common approach for text clustering~\cite{xu2017self, hadifar2019self, zhang2021supporting}. 
(2) \textit{LabelCluster}~\cite{huang2025text} generates candidate labels for records in batches, merges them into $k$ labels, and assigns each record to a label via row-by-row LLM calls. (3) \textit{Lotus}~\cite{patel2024lotus} uses an LLM to predict groups based on the task instruction, then refines them with embedding-based K-means (we use the semantic group-by operator).  (4) \textit{Keyphrase}~\cite{viswanathan-etal-2024-large} augments each row with LLM-generated keyphrases derived from the clustering instruction and applies embedding-based K-means to the augmented rows.

\textbf{Models.} We use OpenAI GPT-4.1 (as $\mathcal{M}_e$) and GPT-4.1-Nano (as $\mathcal{M}_c$) as our LLMs and  text-embedding-3-large as embedding model.

\textbf{Metrics.}
We report cost and accuracy for different methods. For classification and scoring we report accuracy as defined in Eq.~\ref{eq:classification_acc}. For scoring, we additionally consider \textit{pairwise accuracy} that evaluates whether pairs of records are ordered correctly based on their output scores. Given the ground-truth scores $Y = \{y_{i}\}_{i=1}^n$ and predicted scores $\hat{Y} = \{\hat{y}_{i}\}_{i=1}^n$, pairwise accuracy is defined as:
$$
\mathcal{A}_{\text{scr}}(Y, \hat{Y}) = \frac{2}{n(n-1)}\sum_{i=1}^n\sum_{j=i+1}^n \mathds{I}\!\left[\mathrm{sgn}(\hat{y}_{i} - \hat{y}_{j}) = \mathrm{sgn}(y_{i} - y_{j})\right],
$$
where $\mathrm{sgn}(x)$ is the sign function. 
For clustering, given the predicted clusters $\hat{C} = \{\hat{C}_i\}_{i=1}^k$ and ground-truth clusters $C = \{C_j\}_{j=1}^k$, we use \textit{clustering accuracy}, which measures how well the predicted clusters align with the ground-truth clusters, formally:
\vspace{-0.1cm}
$$
\mathcal{A}_{\text{clu}}(C, \hat{C}) = \frac{1}{n}\sum_{i=1}^{k} \max_{j=1}^k \left| \hat{C}_i \cap C_j \right|.\vspace{-0.1cm}
$$
Due to budgetary constraints, we evaluate each method on a random sample subset of 2,000 records for all datasets to estimate accuracy and cost. We report US\$ cost per 1,000 points to normalize cost across datasets irrespective of data size (i.e., for a dataset, $\mathcal{D}$, we report total cost divided by $|\mathcal{D}|$ multiplied by 1,000).

\textbf{Parameters.}
For \sys we use $m_{\max} = 800$, $|S| = 80$ and $B = \max(200, 10 \cdot k)$---we evaluate sensitivity to parameters in \iftoggle{techreport}{our technical report~\cite{techrep}}{Appx.~\ref{appx:exp}} where we observe that a large set of values work well across datasets. By default, we let cost budget $\widetilde{C}=\infty$ (i.e., the goal is to obtain maximum accuracy without a budgetary constraint) but we also present results at different $\widetilde{C}$ values.
For BARGAIN, we set the accuracy target $0.9$ by default. For Batch and Rubric, we set batch size $B = 50$. Unless otherwise stated, for all the methods we use GPT-4.1.

\begin{figure*}
\vspace{-0.8cm}
    \centering
    \includegraphics[width=\textwidth]{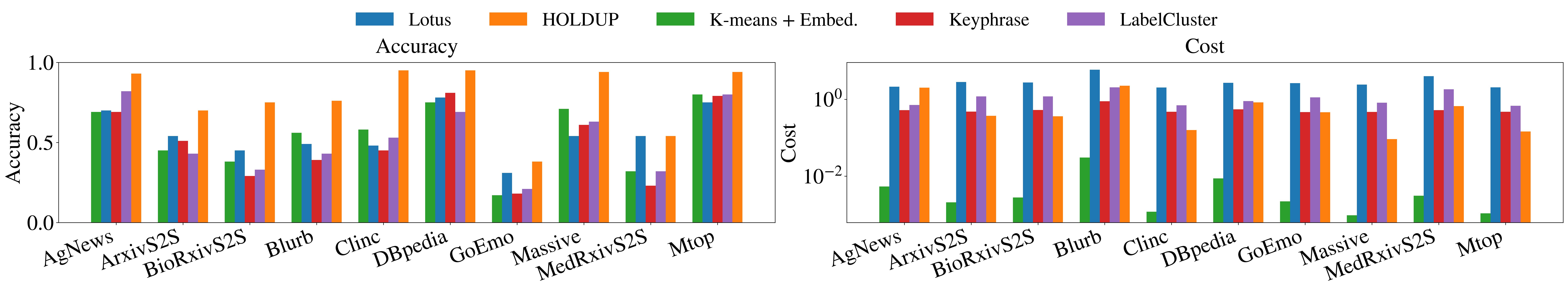}
    \caption{Clustering accuracy and cost across datasets}
    \label{fig:clustering-acc-cost}
\end{figure*}

\subsection{Comparison across Datasets}
\label{sec:experiments:main}

\textbf{Classification.} Fig.~\ref{fig:class-acc} shows the accuracy of \sys compared with various baselines across datasets. \textit{\textbf{\sys consistently improves accuracy over baselines, providing up to 33\% improvement over row-by-row processing.}} Both Batch and Rubric yield worse accuracy than Row-by-Row, as LLM accuracy degrades when processing large data subsets; Rubric is worse because accurately expressing classification criteria in natural language is difficult.

Across datasets, we see that \sys matches Row-by-Row accuracy whenever it is accurate (e.g., on DBPedia and MTop), while often providing significant improvements when it is not. Row-by-Row performs well on datasets with well-defined, distinct classes where the correct label can be inferred without broader data context (e.g., in DBPedia the goal is to classify entities based on type such as ``Artist'' or  ``Athlete''). On the other hand, large accuracy gains for \sys often occurs in datasets with many classes (e.g., ArxivS2S and BiorxivS2S). In such datasets, many class labels are similar and the correct class for a record is ambiguous when it is considered in isolation; \sys resolves such ambiguities by holistically considering the dataset context.

We next evaluate the cost-accuracy trade-offs \sys provides when given a fixed budget on six representative datasets. For BARGAIN, we vary the accuracy target in the set $\{0.1, 0.3, 0.5, 0.7, 0.9\}$ and plot the observed cost and accuracy. 
For \sys, we vary the cost budget, $\widetilde{C}$, to obtain cost comparable to BARGAIN. We drop Rubric from this experiment as it was dominated by Row-by-Row (GPT-4.1) across all datasets in both cost and accuracy.  Fig.~\ref{fig:cost} shows that \sys provides better cost-accuracy trade-offs in general. We observe two categories of datasets. In the first category, \sys improves cost accuracy trade-offs  across all cost regimes (e.g., for ArxivS2S and BiorxivS2S). In fact, for ArxivS2S and BiorxivS2S, \sys is always cheaper than Row-by-Row processing with GPT-4.1, while also providing better accuracy. In the second category, e.g., AGNews and Massive, \sys provides the same cost-accuracy trade-offs as BARGAIN in low-cost regime, but extends the trade-offs to enable more accurate results at higher cost.  In this category, the results show that when given more budget than Row-by-Row processing, we can further improve accuracy with \sys.

\textbf{Scoring}. Fig.~\ref{fig:scoring-acc} shows how \sys compares with baselines for scoring. We see similar accuracy gains as for classification, \textbf{\textit{with \sys consistently improving accuracy over baselines with up to 28\% improvement in accuracy over Row-by-Row processing.}} Trends are similar to classification, although the accuracies, across all methods, are lower for scoring. In general, scoring is a more difficult task, as boundaries between scores are often unclear. 
\sys, by taking data context into account, is able to better understand human preferences and improve scoring accuracy. 

We next evaluate cost-accuracy trade-offs \sys provides at a given cost budget on three representative datasets (results on other datasets were similar). Fig.~\ref{fig:scoring-cost} shows the results.  We see that at the low-cost regime, \sys perform similarly to BARGAIN and Row-by-Row. Indeed, for scoring,  clustering-based classification is often expensive, since the number of iterations, $m$, ends up being large (see Prop.~\ref{prop:cost}). Thus, when cost budget is low, \sys uses row-by-row processing for most records, yielding accuracy similar to BARGAIN and row-by-row processing. However, \sys is able to extend the Pareto frontier,  providing significantly higher accuracy when higher budget is available. For Ticket Support, we see that Batch and Rubric provide better cost-accuracy trade-offs than Row-by-Row because data context is particularly important in this dataset, which also causes those approaches to outperform \sys since it mostly uses Row-by-Row at low budget regime. Indeed, changing \sys to use Batch and Rubric as proxies in the cascade architecture can enable it to also benefit from their advantages, which we leave to future work.  

\textbf{Clustering}. Fig. ~\ref{fig:clustering-acc-cost} compares \sys with baselines for clustering. 
Results are similar to other tasks, with \textbf{\textit{\sys consistently providing better accuracy, outperforming baselines by up to 30\%}}. \sys also provides lower cost compared with baselines that use LLMs. 
These results show that simple ways of directly using LLMs for clustering, or combining them with embedding-based methods, lead to poor accuracy and high cost. Our approach uses LLMs judiciously, achieving better accuracy at lower cost.

\begin{table}[t]
\centering
\footnotesize
\caption{Impact of edge weight calculation on classification}
\label{tab:label-quality}
\setlength{\tabcolsep}{3pt}
\begin{tabular}{lcccccc}
\toprule
\multirow{2}{*}{Method} & \multicolumn{2}{c}{AgNews} & \multicolumn{2}{c}{BiorxivS2S} & \multicolumn{2}{c}{Clinc} \\
\cmidrule(lr){2-3}\cmidrule(lr){4-5}\cmidrule(lr){6-7}
& Acc & Cost & Acc & Cost & Acc & Cost \\
\midrule
\sys with cluster labels & 0.33 & 0.24 & 0.18 & 0.19 & 0.24 & 0.36 \\
\sys with pairwise labels & 0.72 & 39.75 & 0.45 & 36.83 & 0.32 & 29.33\\
\sys minus transitivity & 0.85 & 1.85 & 0.71 & 0.28 & \textbf{0.87} & 0.27\\
\sys & \textbf{0.97} & 2.00 & \textbf{0.81} & 0.36 & \textbf{0.87} & 0.16 \\
\bottomrule
\end{tabular}
\end{table}

\vspace{-0.2cm}
\subsection{Ablation Studies}
\label{sec:expreriments:ablation}
We study two key design choices in \sys: how LLM annotations are obtained to compute edge weights, and how clusters are assigned to classes and scores. We use a subset of representative datasets in this sections---results on other datasets were similar. 




\textbf{Computing Edge Weights.}
We study the importance of our approach for obtaining edge weights in our clustering algorithm (Alg.~\ref{alg:semantic-cluster})---recall that we iteratively obtain edge weights by querying an LLM on subsets of the dataset to obtain record pairs that belong to the same class and take their transitive closure (lines 1-3 in Alg.~\ref{alg:sampling}). We compare this approach with the following methods. We only modify how edge weights are computed, keeping the rest of the correlation clustering algorithm the same. \textit{\sys minus transitivity} is the same as \sys but does not include transitive closure. \textit{\sys with cluster labels} modifies how the LLM is queried: In Alg.~\ref{alg:sampling}, it asks the LLM to output a clustering of the input data, $S_i$ (instead of asking it to output pairs that belong to the same class), then generates the positive annotation set $\bar{P}_i$ as the pairs that belong to the same output cluster (edge weight computation based on $\bar{P}_i$ is kept the same). \textit{\sys with pairwise labels} performs a single LLM call for every pair of records, asking whether they belong to the same class without additional data context, assigning the binary LLM output as the pair's edge weight.

Table~\ref{tab:label-quality} shows how accuracy and cost change based on edge weights. All ablations perform worse than \sys, justifying our approach for computing edge weights. Surprisingly, we see that asking an LLM to output a clustering of the records (i.e., \sys with cluster labels) yields much worse accuracy---we suspect because LLMs are not good at directly clustering a large set of input data due to long context issues. Using pairwise labels (i.e., \sys with pairwise labels) also yields worse accuracy and higher cost, as labels are not created with a holistic data understand. Finally, the results also justify the use of transitive closure. 

\begin{table}[t]
\vspace{-0.82cm}
\centering
\footnotesize
\caption{Comparison of cluster assignment algorithms}

\begin{subtable}[t]{0.48\textwidth}
\setlength{\tabcolsep}{2pt}
\centering
\caption{Classification}
\vspace{-0.2cm}
\label{tab:assignment-class}
\begin{tabular}{lcccc}
\toprule
Method & AgNews & BiorxivS2S & ArxivS2S\\
\midrule
Independent Assignment   & \textbf{0.97} & \textbf{0.81} & 0.55\\
Assignment with Examples &\textbf{0.97}& 0.78 & 0.51\\
Majority Assignment      &\textbf{0.97}& 0.59  & 0.48\\
\sys (Bipartite matching) & \textbf{0.97} &\textbf{0.81} & \textbf{0.75}\\
\bottomrule
\end{tabular}
\end{subtable}
\hfill
\begin{subtable}[t]{0.48\textwidth}
\centering
\caption{Scoring}
\vspace{-0.2cm}
\label{tab:assignment-scoring}
\begin{tabular}{lcccccc}
\toprule
\multirow{2}{*}{Method} & \multicolumn{2}{c}{Yelp} & \multicolumn{2}{c}{Sentiment} & \multicolumn{2}{c}{Ticket Support}\\
\cmidrule(lr){2-3} \cmidrule(lr){4-5} \cmidrule(lr){6-7}
& $\mathcal{A}$ & $\mathcal{A}_{\text{scr}}$ & $\mathcal{A}$ & $\mathcal{A}_{\text{scr}}$ & $\mathcal{A}$ & $\mathcal{A}_{\text{scr}}$ \\
\midrule
Independent Assignment & 0.50 & 0.62 &\textbf{0.87} & \textbf{0.92 }& 0.29 & 0.25\\
Bipartite Matching & \textbf{0.78} &\textbf{0.77} & 0.55 & 0.41 & 0.26 & 0.25 \\
\sys (ILP) & \textbf{0.78} &\textbf{0.77}& \textbf{0.87} &\textbf{0.92} & \textbf{0.42} & \textbf{0.53} \\
\bottomrule
\end{tabular}
\end{subtable}

\end{table}

\textbf{Cluster Assignment Algorithm}.
We next discuss the choice of cluster assignment algorithm for classification and scoring.

Recall that our classification method uses bipartite matching to assign classes to clusters (Alg.~\ref{alg:classification-match}). We compare this choice against three alternatives: \textit{Independent Assignment}, which asks an LLM to independently assign each cluster to a class based on the cluster's content alone; \textit{Assignment with Examples}, which additionally provides example records from other clusters; and \textit{Majority Assignment}, which classifies each record within a cluster independently and assigns the cluster to the majority prediction.


Table~\ref{tab:assignment-class} shows that \sys, which uses bipartite matching, outperforms other alternatives. The benefits are dataset dependent. For AgNews which has few classes, all methods perform well. On BiorxivS2S majority assignment performs poorly, but others remain competitive. Finally, on ArxivS2S, only bipartite matching achieves high accuracy. Indeed, ArxivS2S contains many similar classes, causing methods that independently assign each cluster to make mistakes; bipartite matching avoids this by jointly considering all clusters and classes simultaneously.

Next, we compare our ILP-based cluster assignment (i.e., Alg.~\ref{alg:scoring-sort}) for scoring to other alternatives, specifically, using \textit{Independent Assignment} method, discussed above, as well as \textit{Bipartite Matching}, which uses the same algorithm for classification (Alg.~\ref{sec:matching}) also for scoring. 
Table~\ref{tab:assignment-scoring} shows the result of this comparison, we see that the ILP based method performs best across dataset. 

\vspace{-0.18cm}
\section{Related Work}
\vspace{-0.03cm}

\textbf{Semantic Data Processing}. A large body of recent work has focus on enabling processing unstructured data based on its semantics. These work broadly fall into two categories: (1) work that develop systems and optimizations for general purpose semantic data processing, such as ~\cite{shankar2024docetl, patel2024lotus, russo2025abacus, liu2024declarative, jo2024thalamusdb, urban2024eleet, lin2024towards, wei2025multi, zeighami2025llm, urban2024demonstrating} and (2) work that focus on optimizations for specific operations or tasks within semantic data processing, such as~\cite{shankar2026task, zeighami2025featurized, arora2023language, lin2025twix, xu2026modora, zeighami2024nudge, jo2025sparellm}. Unlike \sys, none of these work considers data context to improve accuracy when performing operations, all performing map, filter and joins independently for each record, thus yielding poor accuracy for tasks when holistic data understanding is needed. We next discuss existing optimizations for classification, scoring and clustering.   

\textbf{Classification and Scoring}.
Before LLMs, a large body of work in NLP had focused on text classification and scoring, e.g., see ~\cite{kowsari2019text, zhang2015character, kim2014convolutional}.
Recent work has leveraged LLMs for text classification~\cite{sun2023text, patel2024lotus, liu2024declarative, shankar2024docetl, jo2024thalamusdb}, processing each row independently with an LLM. Most recent work focuses on cost optimizations for classification, e.g., using model cascade~\cite{zeighami2025cut, patel2024lotus}, or its generalization, task cascade~\cite{shankar2026task}, as well as other architectures that take advantage of cheaper model~\cite{jo2025sparellm, huang2025thriftllm} to reduce cost. 
DocETL~\cite{shankar2024docetl, wei2025multi} uses query rewriting to improve accuracy, but still performs the operations independently for each row.
As our experiments show, \sys outperforms Row-by-Row processing.

\textbf{Clustering}.
Text clustering is also a traditional NLP tasks, with many solutions using clustering algorithm on text embeddings~\cite{reimers2019sentence, xie2016unsupervised}.
More recent work has combined such embedding-based approaches with LLMs, with~\cite{viswanathan2024large, patel2024lotus} using LLMs to modify input text that is embedded, and ~\cite{pattnaik2024improving, viswanathan2024large, diaz-rodriguez2026summaries} using LLMs to guide internals of embedding-based clustering algorithms. Another line of work has focused on LLM-only solutions~\cite{jo2025zerodl, huang2025text}, treating clustering as a classification task by first asking an LLM to generate descriptions of potential clusters and then assigning records to clusters based on which description they match. \sys only uses LLMs, but unlike prior methods, uses a novel graph clustering approach to cluster records, which includes a novel bagging-like approach to obtain feedback from LLMs to decide the probability that two records belong to the same cluster. Our extensive experiments show that \sys outperforms existing clustering approaches that use LLMs or embedding models. 

\textbf{Crowdsourcing.} Many ideas in this paper are inspired by crowdsourcing literature. The observation that data and human context influences judgments has been used in crowdsourcing~\cite{zhuang2015debiasing, zhuang2015leveraging, scholer2013effect, mozer2010decontaminating} although, unlike our work, these papers aim at reducing such contextual effects when obtaining judgments from humans. Our cluster assignment method for scoring combines pairwise rankings to obtain an ordering of the clusters, a problem that has been studied in crowdsourcing~\cite{zhang2016crowdsourced, guo2012so, dushkin2018top}, with our objective similar to that of Kemeny ranking~\cite{kemeny1959mathematics, conitzer2006improved}. More broadly, designing budget-aware schemes to obtain accurate labels for data processing is reminiscent of crowdsourcing work~\cite{welinder2010online, karger2014budget, sheng2008get, gomes2011crowdclustering, marcus2011human, davidson2013using, parameswaran2023revisiting}. 
However, LLMs differ substantially from human annotators in when they provide accurate labels, how those labels can be leveraged for classification, scoring, and clustering, and what cost–accuracy trade-offs they offer~\cite{parameswaran2023revisiting}. These differences necessitate our novel algorithms for querying LLMs and using their responses to perform these tasks.

\vspace{-0.13cm}
\section{Conclusion}
\vspace{-0.03cm}
\label{sec:conclusions}
In this paper, we presented \sys, a method for semantic data processing with holistic data understanding. We identified a key limitation of existing semantic operators: by processing records independently, they rely solely on the LLM's dataset-agnostic interpretation of the task, often leading to inaccurate results. \sys instead collectively processes records and leverages cross-record relationships to ground task interpretation in the data context. Using this insight, we introduced novel clustering, classification and scoring algorithms that process records accurately and at low cost while holistically considering cross-record relationships. Evaluation on 15 real-world datasets demonstrates consistent improvements over existing approaches, with accuracy gains of up to 33\% for classification and 30\% for scoring and clustering tasks. 

\clearpage
\balance
\bibliographystyle{ACM-Reference-Format}
\bibliography{ref}

\appendix
\section{Overview}
This appendix is organized as follows. 
\begin{itemize}
    \item Appx.~\ref{appx:exp} presents additional details.
    \item Appx.~\ref{appx:ilp} provides our ILP formulation for clustering assignment during scoring.
    \item Appx.~\ref{appx:proofs} presents proofs of our theoretical statements.
\end{itemize}
\section{Additional Experiments and Details}\label{appx:exp}
\subsection{Additional Dataset Details}
For the ground-truth labels columns across datasets, we use the `domain' column for the Clinic dataset, Massive and Mtop dataset, `d1' hierarchy level for Blurb, and  
`primary category' for ArxivS2S, BiorxivS2S, and MedrxivS2S datasets. 
Prompts used for all datasets is available in our code base~\cite{holdup_code}.

\begin{figure}
    \centering
    \includegraphics[width=\columnwidth]{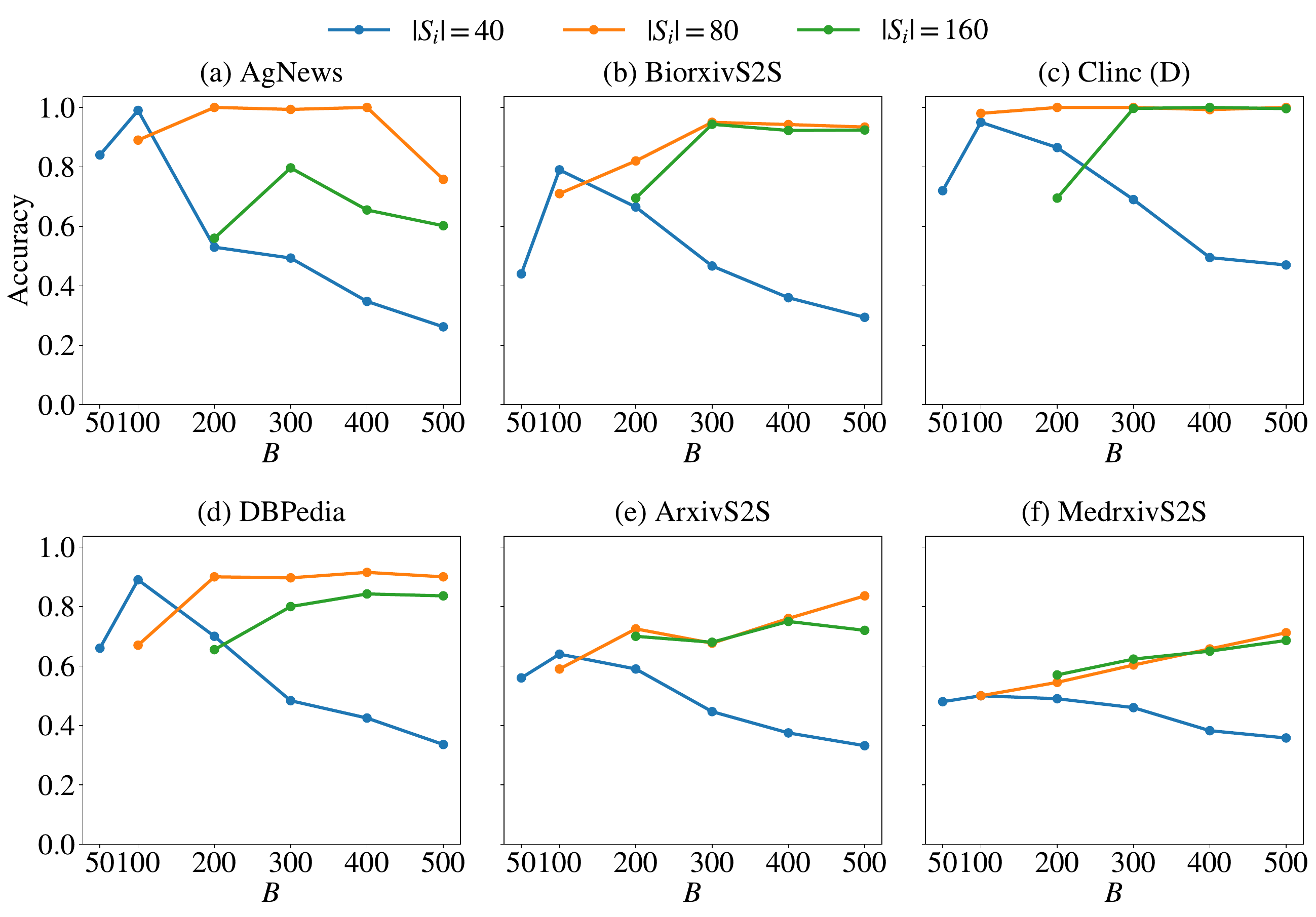}
    \caption{Effect of $m$ and $|S_i|$.}
    \label{fig:batch-sample}
\end{figure}

\subsection{Sensitivity to Parameters}\label{sec:exp:sensitiviy}
We evaluate the effect of $B$ and $|S_i|$ jointly across the six datasets, where $B$ is the number of records clustered each time (i.e., size of input to clustering algorithm~\ref{sec:clustering}) and $S_i$ is size of the sample set passed to the LLM to obtain labels from (i.e., in line 2 of Alg.~\ref{alg:sampling}). We let $m_{\max}=2,000$ which caps the total number of samples taken (i.e., in line 2 of Alg.~\ref{alg:semantic-cluster}).
Figure ~\ref{fig:batch-sample} demonstrates how clustering accuracy changes based on $B$ and $|S_i|$.
$|S_i| = 80$ in general achieves the highest accuracy compared to $|S_i| = 40$ and $|S_i| = 160$, although $|S_i| = 80$ and $|S_i| = 160$ are comparable in many datasets. Regarding $B$, we see that accuracy generally increases as $B$ increases, except for $|S_i|=40$. For $|S_i| = 40$, the accuracy is in general lower and decreases as $B$ increases. This is due to two factors. First, a small $|S_i|$ means each LLM call provides fewer annotations, so that more LLM calls would be necessary to obtain enough annotations (number of LLM calls are currently capped at 80). Second,  $|S_i|$ might not be insufficient to provide data context, especially for datasets with many clusters (e.g., if the number of clusters is more than 40). 
$|S_i|=160$ often performs slightly worse that $|S_i|=80$, we suspect because the longer context causes more LLM error.

\begin{table}[]
    \begin{subtable}[t]{\columnwidth}
\begin{adjustbox}{width=\textwidth}
\begin{tabular}{lcccccc}
\toprule
\multirow{2}{*}{Method} & \multicolumn{2}{c}{AgNews} & \multicolumn{2}{c}{BiorxivS2S} & \multicolumn{2}{c}{Clinc} \\
\cmidrule(lr){2-3}\cmidrule(lr){4-5}\cmidrule(lr){6-7}
& Acc & Cost & Acc & Cost & Acc & Cost \\
\midrule
\sys with clustering labels & 0.26 & 0.33 & 0.10 & 0.05 & 0.19 & 0.31 \\
\sys with pairwise labels & 0.72 & 39.83 & 0.14 & 40.29 & 0.30 & 29.37\\
\sys minus transitive closure & 0.85 & 1.76 & 0.73 & 0.14 & 0.87 & 0.22 \\
\sys & \textbf{0.97} & 1.97 & \textbf{0.83} & 0.32 & \textbf{0.87} & 1.68 \\
\bottomrule
\end{tabular}
\end{adjustbox}
\caption{Clustering}
\label{tab:label-quality-clustering}
\end{subtable}
\end{table}
\subsection{Ablation of Edge Weights for Clustering}
We present an ablation study analogous to Sec.~\ref{sec:expreriments:ablation}, where we evaluate impact of ablations of \sys on clustering accuracy. The baselines we consider here are as discussed in Sec.~\ref{sec:expreriments:ablation}. The results are shown in Table~\ref{tab:label-quality-clustering} where we see trends similar to  Sec.~\ref{sec:expreriments:ablation}.
\section{ILP Details for Scoring}\label{appx:ilp}

We formulate an ILP to assign each cluster a unique score to minimize the total violations of LLM's pairwise comparisons.
We define $a_{i,j} \in \{0, 1\}$ to be an indicator variable denoting whether group $\hat{C}_i$ is assigned to score $j$.
We define $b_{i, j} \in \{0, 1\}$ to be an indicator variable denoting $C_i$ is assigned a higher score than $\hat{C}_j$.
That is, $b_{i, j} = \mathds{I}[\pi(\hat{C}_i) > \pi(\hat{C}_j)]$.
Finally, we build the objective and the constraints as follows:
$$
\renewcommand{\arraystretch}{1.4}
\begin{array}{lll}
\min \quad  & \multicolumn{2}{l}{\sum_{1 \leq i < j \leq k} \left( b_{i,j} \cdot W(\hat{C}_i, \hat{C}_j) + (1 - b_{i,j}) \cdot W(\hat{C}_j, \hat{C}_i)\right)} \\
\text{s.t.}\quad & \sum_{s=1}^{k} a_{i,s} = 1 & \forall i = 1\dots,k \quad \text{(score uniqueness)} \\
& \sum_{i=1}^{k} a_{i,s} = 1 & \forall s = 1,\dots,k \quad \text{(score distinctness)} \\
& \pi(\hat{C}_i) = \sum_{s=1}^{k} s\cdot a_{i,s} & \forall i\quad \text{(score assignment)} \\
&b_{i,j} = \mathds{I}[\pi(\hat{C}_i) > \pi(\hat{C}_j)] & \forall i < j\quad \text{(pairwise relationship)} \\
& a_{i,s} \in \{0,1\} & \forall i,s \\
\end{array} 
$$
Note that $b_{i,j} = \mathds{I}[\pi(\hat{C}_i) > \pi(\hat{C}_j)]$ can be implemented in an ILP with a big-M constraint. By solving the ILP problem, we obtain an optimal scoring of the clusters that minimizes the violation of pairwise comparison results.
\section{Proofs}\label{appx:proofs}
\subsection{Proof of Lemma~\ref{lemma:decidability}}
We use a simplified statistical model of the sampling procedure. We first discuss necessary notation, then define our statistical model and its assumption before deriving Lemma~\ref{lemma:decidability}.

Let $E_{z}^{a, b}$ be a Bernoulli random variable denoting the sampling outcome when the edge between $t_a$ and $t_b$ is sampled, so that $W[a, b]=\frac{1}{r}\sum_{z\in [r]}E_z^{a, b}.$ Recall that 
\begin{align*}
d(a, j) &= \sum_{\overset{id_b = j,}{a \neq b}} W[a, b] + \sum_{\overset{id_b \neq j,}{a \neq b}} \left(1 - W[a, b]\right)
\end{align*}
so that 
\begin{align*}
d(a, j) &= \frac{1}{r}\sum_{\overset{id_b = j,}{a \neq b}} \sum_{z\in [r]}E_z^{a, b} + \frac{1}{r}\sum_{\overset{id_b \neq j,}{a \neq b}} \left(\sum_{z\in [r]}(1- E_z^{a, b})\right)\\
&= \frac{1}{r}\sum_{z\in [r]}\sum_{\overset{id_b = j,}{a \neq b}} E_z^{a, b} + \frac{1}{r}\sum_{z\in [r]}\sum_{\overset{id_b \neq j,}{a \neq b}} \left(1 - E_z^{a, b}\right)
\end{align*}
Now denote by $D^{a, j}_z=\sum_{\overset{id_b = j,}{a \neq b}} E_z^{a, b} + \sum_{\overset{id_b \neq j,}{a \neq b}} (1 - E_z^{a, b})$ for $z\in [r]$. 

We use a statistical model with the following assumptions. First, we assume  $D^{a, j}_1, ..., D^{a, j}_r$ are i.i.d random variables with true mean $d^*(a, j)$ and empirical mean $d(a, j)$. Furthermore, we assume $D^{a, j}_z$ are independently computed for each $z$ and $j$  so that all $D^{a, j}_z$ are independent from each other. This assumption models a statistical process where new edge weights are independently sampled to estimate each $D^{a, j}_z$. We also make an assumption that $D^{a, j}_z\leq|\hat{C}_j|$, that is, the disagreement for a record is bounded by the size of the cluster, which we expect to  be true in practice. 

Based on these assumptions, then
\begin{align*}
\mathds{P}(d(a, i)-d^*(a, i)\geq\epsilon_a,& i\in [k]\setminus\{id_a\})\\&=\Pi_{i\in [k]\setminus\{id_a\}}\mathds{P}(|d(a, i)-d^*(a, i)|\geq\epsilon_a)\\
&\leq\Pi_{i\in [k]\setminus\{id_a\}}\exp{(2r\frac{\epsilon_a^2}{|C_i|^2})}\\
&=\exp{(2r\sum_{i\in [k]\setminus\{id_a\}}\frac{\epsilon_a^2}{|C_i|^2}).}
\end{align*}
Similarly 
\begin{align*}
\mathds{P}(d^*(a, id_a)-d(a, id_a)\geq\epsilon_a)&\leq \exp{(2r\frac{\epsilon_a^2}{|C_{id_a}|^2}).}
\end{align*}
Thus, combining the above, 

\begin{align*}
\mathds{P}(d(a, i)-d^*(a, i)\geq\epsilon_a, i\in [k]\setminus\{id_a\},d^*(a, id_a)-d(a, id_a)\geq\epsilon_a)\\
\leq \exp{(2r\sum_{i\in [k]}\frac{\epsilon_a^2}{|C_i|^2}).}
\end{align*}

\subsection{Proof of Prop~\ref{prop:cost}}
The total cost is divide into cost of performing \texttt{CbClassification} and row-by-row processing. We perform \texttt{CbClassification} on two subsets of $\mathcal{D}$: the sample batch $D_0$ (lines 1-2,  Alg.~\ref{alg:classification-framework}) and the subset $D_X \subseteq \mathcal{D} \setminus D_0$  that are above the confidence threshold (lines 4-5,  Alg.~\ref{alg:classification-framework}). We perform row-by-row classification on all records in $\mathcal{D} \setminus D_0$ (line 3, Alg.~\ref{alg:classification-framework}). We next discuss the cost of row-by-row classification and  \texttt{CbClassification} separately.

\textit{Row-by-row Classification}. For row-by-row classification, each LLM call to the chosen proxy, $\mathcal{M}^P$ contains a single record $t$ and all class names $\ell_1, \cdots, \ell_k$ (Alg.~\ref{alg:cascading}, lines 9-10). Let $c_P$ be the cost per token for proxy, $\mathcal{M}^P$. Then, cost of this step is $$c_P \cdot |\mathcal{D}\setminus D_0|\cdot (l_r + L_\ell) = c_P \cdot (n - B)\cdot (l_r + L_\ell)\leq c_P\cdot ( L_r + n\cdot L_\ell).$$

\textit{\texttt{CbClassification}}. This step consists of clustering records and cluster assignment, applied to $\left(1 + \lceil\frac{|D_X|}{B}\rceil\right) = 1 + n_B$, batches, where $n_B=\lceil\frac{|D_X|}{B}\rceil$. Let the LLMs used in clustering records and cluster assignment be $\mathcal{M}^{\text{cluster}}$, and $\mathcal{M}^{\text{assign}}$, with unit prices $c_{\text{cluster}}, c_{\text{assign}}$ respectively.

For clustering, each call to Alg.~\ref{alg:semantic-cluster} processes a batch, $D_B$. The cost for each call is due to $m$ LLM calls $\mathcal{M}^{\text{cluster}}(\texttt{P}, S_i)$ on sampled subsets $S_i\subseteq D_B, i \in [m]$ (Alg.~\ref{alg:sampling}, line 3), where $m$ is to total number of iterations (i.e., for loop in line 2 in Alg.~\ref{alg:sampling}). Each call includes $|S_i|$ records, resulting in a prompt length of $|S_i| \cdot l_r$. The total clustering cost is: $c_\text{cluster} \cdot m \cdot |S_i| \cdot l_r$.

For cluster assignment (i.e., Alg.~\ref{alg:classification-match}) on a batch $D_B$ clustered into sets $\hat{C}_1, ...,  \hat{C}_k$, the cost is due to $k$ LLM calls $\mathcal{M}^{\text{assign}}(\texttt{P}, \hat{C}_i, \ell)$, one for each cluster $\hat{C}_i$, to evaluate its alignment with all $k$ class names (line 3). Each call includes $|\hat{C}_i|$ records and all $k$ class names. The cost is:
$$
c_{\text{assignment}}\cdot \sum_{i=1}^k \left(|\hat{C}_i|\cdot l_r + k \cdot l_\ell\right)
= c_{\text{assignment}}\cdot (B \cdot l_r + k \cdot L_\ell)
$$

\textit{Total Cost.} As \texttt{CbClassification} is applied to a total of $\left(1 + n_B\right)$ batches on $D_0\cup D_X$, the total cost is at most:
$$
\varkappa(c_P  (L_r + nL_\ell) + 
c_{\text{cluster}} \cdot m  \cdot r \cdot L_r + c_{\text{assignment}}\cdot r\cdot (L_r + n\cdot k \cdot L_\ell)) 
$$
where $\varkappa$ is a constant. Rearenging gives the statment of the proposition. 

\end{document}